\documentclass[a4]{article}
\usepackage[margin=10em]{geometry}
\usepackage{multirow}
\usepackage{amsmath, array, booktabs, caption, fancyvrb, graphicx, hypdoc, libertine, longtable, newtxmath, tabularx, zi4}
\usepackage{authblk}
\usepackage{listings}
\usepackage{natbib}
\usepackage{subcaption}
\usepackage[T1]{fontenc}
\usepackage[inline,shortlabels]{enumitem}
\usepackage{fixfoot}

\begin{document}

\newcommand{\Danon}[2]{#1}
\newcommand{\Dodona}[1]{\Danon{Dodona}{Naos}}
\newcommand{\Durl}[1]{\Danon{\url{#1}}{masked URL}}

\title{\Dodona{}: learn to code with a virtual co-teacher that supports active learning}

\author{{Charlotte} {Van Petegem}}
\affil{charlotte.vanpetegem@ugent.be}

\author{Rien Maertens}
\affil{rien.maertens@ugent.be}

\author{Niko Strijbol}
\affil{niko.strijbol@ugent.be}

\author{{Jorg} {Van Renterghem}}
\affil{jorg.vanrenterghem@ugent.be}

\author{{Felix} {Van der Jeugt}}
\affil{felix.vanderjeugt@ugent.be}

\author{{Bram} {De Wever}}
\affil{bram.dewever@ugent.be}

\author{Peter Dawyndt}
\affil{peter.dawyndt@ugent.be}

\author{Bart Mesuere}
\affil{bart.mesuere@ugent.be}

%
\begin{abstract}
  \Dodona{} (\Durl{dodona.ugent.be}) is an intelligent tutoring system for
  computer programming. It bridges the gap between assessment and learning by
  providing real-time data and feedback to help students learn better, teachers
  teach better and educational technology become more effective. We demonstrate
  how \Dodona{} can be used as a virtual co-teacher to stimulate active learning
  and support challenge-based education in open and collaborative learning
  environments. We also highlight some of the opportunities (automated feedback,
  learning analytics, educational data mining) and challenges (scalable
  feedback, open internet exams, plagiarism) we faced in practice. \Dodona{} is
  free for use and has more than 36 thousand registered users across many educational
  and research institutes, of which 15 thousand new users registered last year. Lowering
  the barriers for such a broad adoption was achieved by following best
  practices and extensible approaches for software development, authentication,
  content management, assessment, security and interoperability, and by adopting
  a holistic view on computer-assisted learning and teaching that spans all
  aspects of managing courses that involve programming assignments. The source
  code of \Dodona{} is available on GitHub under the permissive MIT open-source
  license.
\end{abstract}

\maketitle

%
\section*{Keywords}
computer programming, education, active learning, intelligent tutoring system, computer-assisted learning, computer-assisted teaching, classroom management, decentralized authentication, automated assessment, feedback, plagiarism, learning analytics, educational data mining

%

\Danon{}{\textbf{Author's Note:} The name of our platform has been replaced with
  a made-up alternative. Links to documentation and courses have been replaced
  with a mention that they are masked.}

\section{Introduction}

The only way to learn how to solve problems with computer programs is by solving
lots of problems, and programming assignments are the main way in which such
practice is generated \citep{gibbs_conditions_2005}. Because of its potential to
establish feedback loops that are scalable and responsive enough for an active
learning environment, automated source code assessment has become a driving
force in computer science, statistics and data science courses
\citep{ala-mutka_survey_2005,douce_automatic_2005,ihantola_review_2010,paiva_automated_2022}. Automated
assessment was introduced in programming education in the early 1960s
\citep{hollingsworth_automatic_1960} and enables students to receive immediate
and customized feedback upon each submitted solution without the need for any
human intervention, especially when provided through interactive web
applications \citep{wasik_survey_2018}. Due to the iterative nature of software
development --- and problem solving in general --- establishing such a
collaborative dialogue while students work towards acceptable solutions for
their programming assignments could hardly be achieved by human assessment alone
\citep{bell_connectivism_2011,ihantola_review_2010}. In fact,
\citet{cheang_automated_2003} identified the labor-intensiveness of assessing
programming assignments as the main reason why few such assignments are given to
the students, when ideally they should be given many more. Freeing teachers from
the assessment burden creates possibilities for stimulating students to practice
more often, which is recognized as a good practice for improving programming
competences \citep{woit_effectiveness_2003}.

Given the clear advantages in speed, availability, consistency and objectivity
\citep{ala-mutka_survey_2005}, automated assessment is not only capable of
supporting assessment \textit{of} learning, but also assessment \textit{for}
learning. Assessment \textit{for} learning focuses on formative assessment to
provide feedback to students to enhance their learning and not only summatively
assess their learning at the end. However, when implementing automated
assessment, there is a need for careful pedagogical design of programming
assignments and assessment strategies \citep{forisek_suitability_2006}. Failing
to do so may reduce motivation and stimulate cheating
\citep{wootton_encouraging_2002}. Given the limitations of assessment tools and
depending on a course's learning goals, teaching context and available human
resources, teachers therefore need to decide what feedback they want and can
provide to their students and when, how and by whom it is provided
\citep{gibbs_conditions_2005}. Generic feedback on programming assignments or
customized feedback on student submissions may come as tips \& tricks, code
quality metrics, software testing reports, source code annotations, model
solutions and grades \citep{ala-mutka_survey_2005}. Feedback might be provided
before students start working on an assignment (``feed up''), while they are
working on their solution (``feed back'' on how they performed and ``feed
forward'' on what to do next) or after the submission deadline has passed
(``feed back'' on their overall performance) \citep{hattie_power_2007}. Feedback
that takes some time to generate can only be supplied asynchronously, while
synchronous delivery also becomes an option when feedback is immediately
available \citep{chickering_seven_1987}.

As a result, a lot of pedagogical decisions need to be made when designing and
running courses that involve automatically assessed programming
assignments. What programming language is used \citep{crick_analysis_2017}?
Should introductory programming merely focus on learning the syntax and
semantics of a programming language? Or should it simultaneously also stress the
importance of programming skills for problem solving with computers, readability
through good programming style \citep{rogers_acce_2014}, performance through
data structures and algorithms, or maintainability through writing software
tests \citep{edwards_using_2004,marrero_testing_2005}? Should assessment
primarily report on shortcomings of student submissions or also hint on how
these defects might be remedied \citep{rivers_data-driven_2017}? Should grades
and more elaborate feedback be generated purely based on automated assessment,
or does it also require human intervention
\citep{douce_automatic_2005,jackson_semi-automated_2000}? Should assignments be
restricted to what can reasonably be assessed automatically or should they
remain as authentic as possible? Should students learn to use standard software
development tools for writing, building, running, testing and debugging software
as supplied by modern integrated development environments or should they use
tools specifically designed for use in an educational context? Should students
learn to understand the diagnostic messages generated by compilers and
interpreters or do we provide help deciphering them
\citep{becker_compiler_2019}?  Should students be somehow restricted in
submitting solutions for programming assignments or should feedback from
automated assessment be reduced \citep{ihantola_review_2010}? Without any
exception, all intelligent tutoring systems for automated source code assessment
that were developed over the years have hardcoded some of these choices, either
by explicit design or silently by not supporting
alternatives. \citet{ihantola_review_2010} identified lack of flexibility due to
such hardcoded restrictions as one of the main reasons why few systems have seen
adoption beyond the course or institute where they were initially developed, and
found their open source policy, lifespan, interoperability and portability quite
disappointing.

This paper introduces \Dodona{} (\Durl{dodona.ugent.be}) as an online learning
environment that embraces the importance of active learning and just-in-time
feedback in courses involving programming assignments. After presenting some of
its key features for computer-assisted learning and teaching
(Section~\ref{features}), we discuss how we use the platform in managing an
introductory programming course with a strong focus on active and online
learning (Section~\ref{managing}). This case study can be read as an inspiration
for running programming courses in an open and collaborative learning
environment, but also explains the context that guided some of the design
choices we made in developing \Dodona{}. We further discuss how \Dodona{}
succeeds in breaking the walls for EdTech tools and open educational resources
to be effectively used beyond the context in which they were initially created
(Section~\ref{perspective}).

\section{Key features of \Dodona{}}\label{features}

\Dodona{} is an intelligent tutoring system for computer programming that is
built around a generic infrastructure for automatic assessment and a distributed
model for developing and publishing learning material. This allows it to cope
with the multifaceted nature of assessing source code submitted for programming
assignments by supporting different programming languages, runtime environments,
evaluation criteria, software testing techniques and target audiences. But the
platform also endorses the blended learning idea that feedback can be provided
as a multi-step process, mixing the complementary strengths of learning from
self, peers, instructors, teachers and software agents to deal with their
limitations in producing the required volume and thoroughness
\citep{cooper_facilitating_2000}. In that vision, automated assessment is only a
first step in providing remedial feedback in small chunks while students work on
their programming assignments. The frequency and responsiveness of automated
assessment, which can be provided reasonably economically, compensates for the
basic level and lack of individualization of its feedback
\citep{gibbs_conditions_2005}. To further deal with the economy of scale, the
feedback from automated assessment is used as a stepping stone to streamline
human interventions whenever students ask for more customized feedback or when
reviewing and grading source code submitted during high-stake tests and
exams. \Dodona{} also lowers the barriers for broader adoption of the tool by
following best practices and extensible models for authentication, content
management, assessment, security and interoperability, and by adopting a
holistic view on computer-assisted learning and teaching that spans all aspects
of managing courses, from internationalization and localization to learning
analytics and educational data mining. In what follows, we cover each of these
features in more detail.

\subsection{Classroom management}

In \Dodona{}, a \textbf{course} is where teachers and instructors effectively
manage a learning environment by instructing, monitoring and evaluating their
students and interacting with them, either individually or as a group. A
\Dodona{} user who created a course becomes its first administrator and can
promote other registered users as \textbf{course administrators}. In what
follows, we will also use the generic term teacher as a synonym for course
administrators if this \Dodona{}-specific interpretation is clear from the
context, but keep in mind that courses may have multiple administrators.

\begin{figure*}
  \centering
  \includegraphics[width=\linewidth]{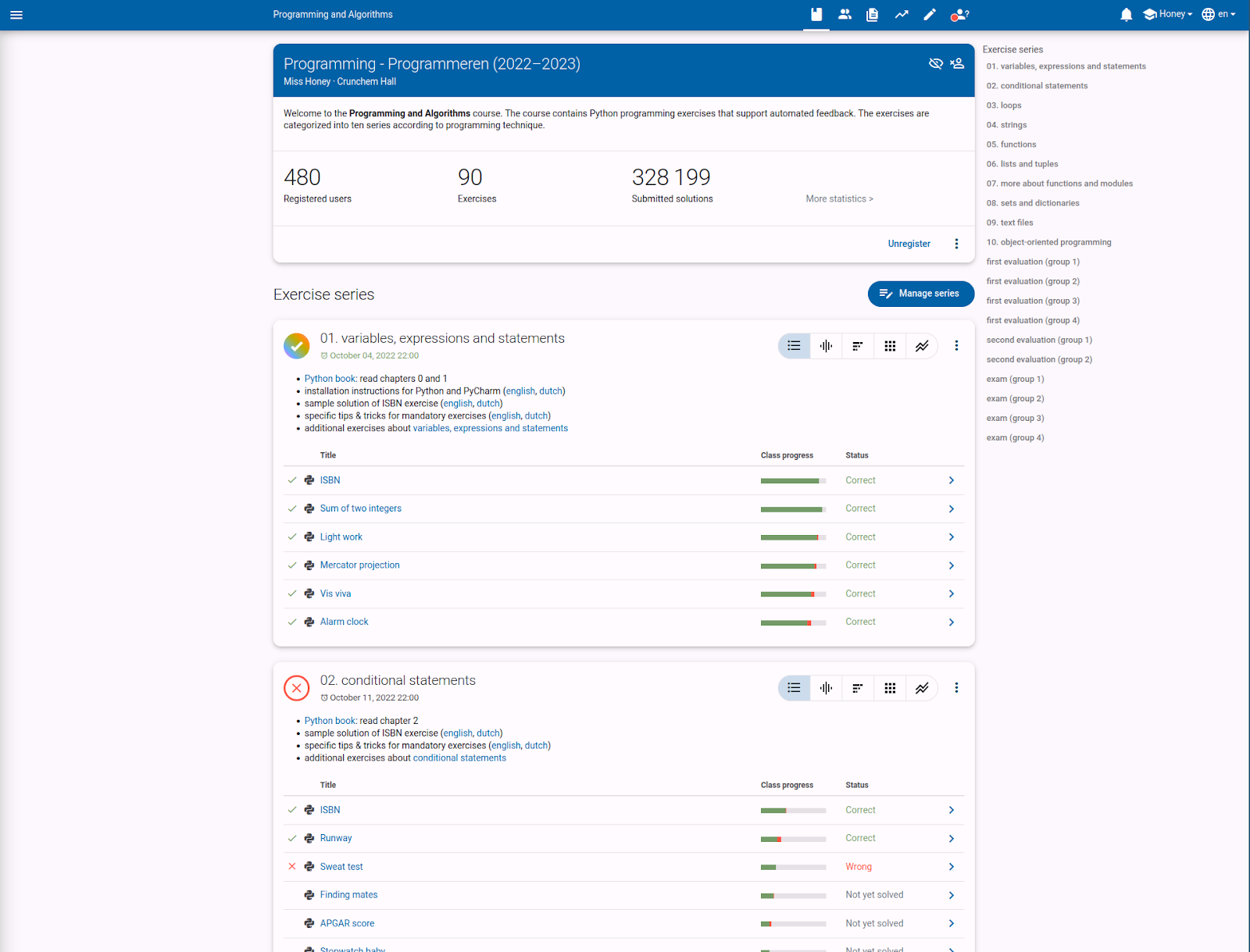}
  \caption{Main course page (administrator view) showing some series with
    deadlines, reading activities and programming assignments in its learning
    path. At any point in time, students can see their own progress through the
    learning path of the course. Teachers have some additional icons in the
    navigation bar (top) that lead to an overview of all students and their
    progress, an overview of all submissions for programming assignments,
    general learning analytics about the course, course management and a
    dashboard with questions from students in various stages from being answered
    (Figure~\ref{fig:questions}). The red dot on the latter icon notifies that
    some student questions are still pending.}\label{fig:course}
\end{figure*}

The course itself is laid out as a \textbf{learning path} that consists of
course units called series, each containing a sequence of \textbf{learning
  activities} (Figure~\ref{fig:course}). Among the learning activities we
differentiate between \textbf{reading activities} that can be marked as read and
\textbf{programming assignments} with support for automated assessment of
submitted solutions. Learning paths are composed as a recommended sequence of
learning activities to build knowledge progressively, allowing students to
monitor their own progress at any point in time. Courses can either be created
from scratch or from copying an existing course and making additions, deletions
and rearrangements to its learning path.

Students can \textbf{self-register} to courses in order to avoid unnecessary
user management. A course can either be announced in the public overview of
\Dodona{} for everyone to see, or be limited in visibility to students from a
certain educational institution. Alternatively, students can be invited to a
hidden course by sharing a secret link. Independent of course visibility,
registration for a course can either be open to everyone, restricted to users
from the institution the course is associated with or new registrations can be
disabled altogether. Registrations are either approved automatically or require
explicit approval by a teacher. Registered users can be tagged with one or more
labels to create subgroups that may play a role in learning analytics and
reporting.

Students and teachers more or less see the same course page, except for some
management features and learning analytics that are reserved for
teachers. Teachers can make content in the learning path temporarily
inaccessible and/or invisible to students. Content is typically made
inaccessible when it is still in preparation or if it will be used for
evaluating students during a specific period. A token link can be used to grant
access to invisible content, e.g.\ when taking a test or exam from a subgroup of
students.

Students can only mark reading activities as read once, but there is no
restriction on the number of solutions they can submit for programming
assignments. Submitted solutions are automatically assessed and students receive
immediate feedback as soon as the assessment has completed, usually within a few
seconds. \Dodona{} stores all submissions, along with submission metadata and
generated feedback, such that the submission and feedback history can be
reclaimed at all times. On top of automated assessment, student submissions may
be further assessed and graded manually by a teacher.

Series can have a \textbf{deadline}. Passed deadlines do not prevent students
from marking reading activities or submitting solutions for programming
assignments in their series. However, learning analytics, reports and exports
usually only take into account submissions before the deadline. Because of the
importance of deadlines and to avoid discussions with students about missed
deadlines, series deadlines are not only announced on the course page. The
student's home page highlights upcoming deadlines for individual courses and
across all courses. While working on a programming assignment, students also
start to see a clear warning from ten minutes before a deadline onwards. Courses
also provide an iCalendar link \citep{stenerson_internet_1998} that students can
use to publish course deadlines in their personal calendar application.

Because \Dodona{} logs all student submissions and their metadata, including
feedback and grades from automated and manual assessment, we use that data to
integrate reports and learning analytics in the course page
\citep{ferguson_learning_2012}. We also provide export wizards that enable the
extraction of raw and aggregated data in CSV-format for downstream processing
and educational data mining
\citep{baker_state_2009,romero_educational_2010}. This allows teachers to better
understand student behavior, progress and knowledge, and might give deeper
insight into the underlying factors that contribute to student actions
\citep{ihantola_review_2010}. Understanding, knowledge and insights that can be
used to make informed decisions about courses and their pedagogy, increase
student engagement, and identify at-risk students
\citep{van_petegem_passfail_2022}.

\subsection{User management}

Instead of providing its own authentication and authorization, \Dodona{}
delegates authentication to external identity providers (e.g.\ educational and
research institutions) through SAML \citep{farrell_assertions_2002}, OAuth
\citep{leiba_oauth_2012,hardt_oauth_2012} and OpenID Connect
\citep{sakimura_openid_2014}. This support for \textbf{decentralized
  authentication} allows users to benefit from single sign-on when using their
institutional account across multiple platforms and teachers to trust their
students' identities when taking high-stakes tests and exams in \Dodona{}.

\Dodona{} automatically creates user accounts upon successful authentication and
uses the association with external identity providers to assign an institution
to users. By default, newly created users are assigned a student role. Teachers
and instructors who wish to create content (courses, learning activities and
judges), must first request teacher
rights\footnote{\Durl{https://dodona.ugent.be/rights\_requests/new/}} using a
streamlined form.

\subsection{Automated assessment}

The range of approaches, techniques and tools for software testing that may
underpin assessing the quality of software under test is incredibly
diverse. Static testing directly analyses the syntax, structure and data flow of
source code, whereas dynamic testing involves running the code with a given set
of test cases
\citep{graham_foundations_2021,oberkampf_verification_2010}. Black-box testing
uses test cases that examine functionality exposed to end-users without looking
at the actual source code, whereas white-box testing hooks test cases onto the
internal structure of the code to test specific paths within a single unit,
between units during integration, or between subsystems
\citep{nidhra_black_2012}. So, broadly speaking, there are three levels of
white-box testing: unit testing, integration testing and system testing
\citep{dooley_software_2011,wiegers_creating_1996}. Source code submitted by
students can therefore be verified and validated against a multitude of
criteria: functional completeness and correctness, architectural design,
usability, performance and scalability in terms of speed, concurrency and memory
footprint, security, readability (programming style), maintainability (test
quality) and reliability \citep{staubitz_towards_2015}. This is also reflected
by the fact that a diverse range of metrics for measuring software quality have
come forward, such as cohesion/coupling
\citep{yourdon_structured_1979,stevens_structured_1999}, cyclomatic complexity
\citep{mccabe_complexity_1976} or test coverage \citep{miller_systematic_1963}.

To cope with such a diversity in software testing alternatives, \Dodona{} is
centered around a generic infrastructure for \textbf{programming assignments
that support automated assessment}. Assessment of a student submission for an
assignment comprises three loosely coupled components: a container, a judge and
an assignment-specific assessment configuration.

\begin{figure}
  \centering
  \includegraphics[width=0.5\linewidth]{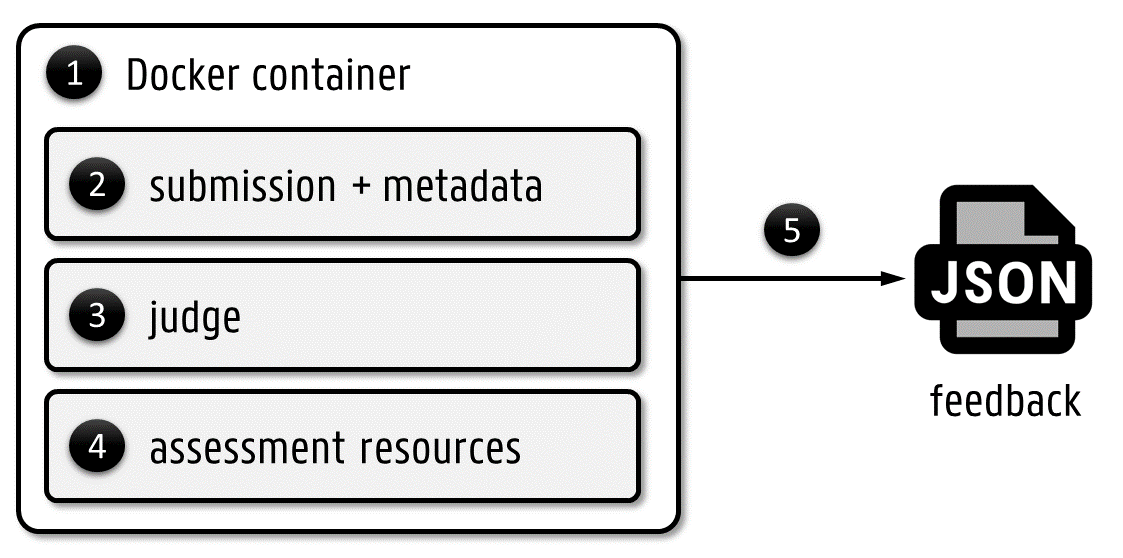}
  \caption{Outline of the procedure to automatically assess a student submission
    for a programming assignment. \Dodona{} instantiates a Docker container (1)
    from the image linked to the assignment (or from the default image linked to
    the judge of the assignment) and loads the submission and its metadata (2),
    the judge linked to the assignment (3) and the assessment resources of the
    assignment (4) into the container. \Dodona{} then launches the actual
    assessment, collects and bundles the generated feedback (5), and stores it
    into a database along with the submission and its metadata.}\label{fig:process}
\end{figure}

For proper virtualization we use \textbf{Docker containers}
\citep{peveler_comparing_2019} that use OS-level containerization technologies
and define runtime environments in which all data and executable software (e.g.,
scripts, compilers, interpreters, linters, database systems) are provided and
executed. These resources are typically pre-installed in the image of the
container. Prior to launching the actual assessment, the container is extended
with the submission, the judge and the resources included in the assessment
configuration (Figure~\ref{fig:process}). Additional resources can be downloaded
and/or installed during the assessment itself, provided that Internet access is
granted to the container.

The actual assessment of the student submission is done by a software component
called a \textbf{judge} \citep{wasik_survey_2018}. The judge must be robust
enough to provide feedback on all possible submissions for the assignment,
especially submissions that are incorrect or deliberately want to tamper with
the automatic assessment procedure \citep{forisek_suitability_2006}. Following
the principles of software reuse, the judge is ideally also a generic framework
that can be used to assess submissions for multiple assignments. This is enabled
by the submission metadata that is passed when calling the judge, which includes
the path to the source code of the submission, the path to the assessment
resources of the assignment and other metadata such as programming language,
natural language, time limit and memory limit.

Rather than providing a fixed set of judges, \Dodona{} adopts a minimalistic
interface that allows third parties to create new
judges\footnote{\Durl{https://docs.dodona.be/en/guides/creating-a-judge/}}:
automatic assessment is bootstrapped by launching the judge's \texttt{run}
executable that can fetch the JSON formatted submission metadata from standard
input and must generate JSON formatted feedback on standard output. The feedback
has a standardized hierarchical structure that is specified in a JSON
schema\footnote{\Durl{https://github.com/dodona-edu/dodona/tree/develop/public/schemas}}. At
the lowest level, \textbf{tests} are a form of structured feedback expressed as
a pair of generated and expected results. They typically test some behavior of
the submitted code against expected behavior. Tests can have a brief description
and snippets of unstructured feedback called messages. Descriptions and messages
can be formatted as plain text, HTML (including images), Markdown, or source
code. Tests can be grouped into \textbf{test cases}, which in turn can be
grouped into \textbf{contexts} and eventually into \textbf{tabs}. All these
hierarchical levels can have descriptions and messages of their own and serve no
other purpose than visually grouping tests in the user interface. At the top
level, a submission has a fine-grained status that reflects the overall
assessment of the submission: \texttt{compilation error} (the submitted code did
not compile), \texttt{runtime error} (executing the submitted code failed during
assessment), \texttt{memory limit exceeded} (memory limit was exceeded during
assessment), \texttt{time limit exceeded} (assessment did not complete within
the given time), \texttt{output limit exceeded} (too much output was generated
during assessment), \texttt{wrong} (assessment completed but not all strict
requirements were fulfilled), or \texttt{correct} (assessment completed and all
strict requirements were fulfilled).

\begin{figure*}
  \centering
  \includegraphics[width=\linewidth]{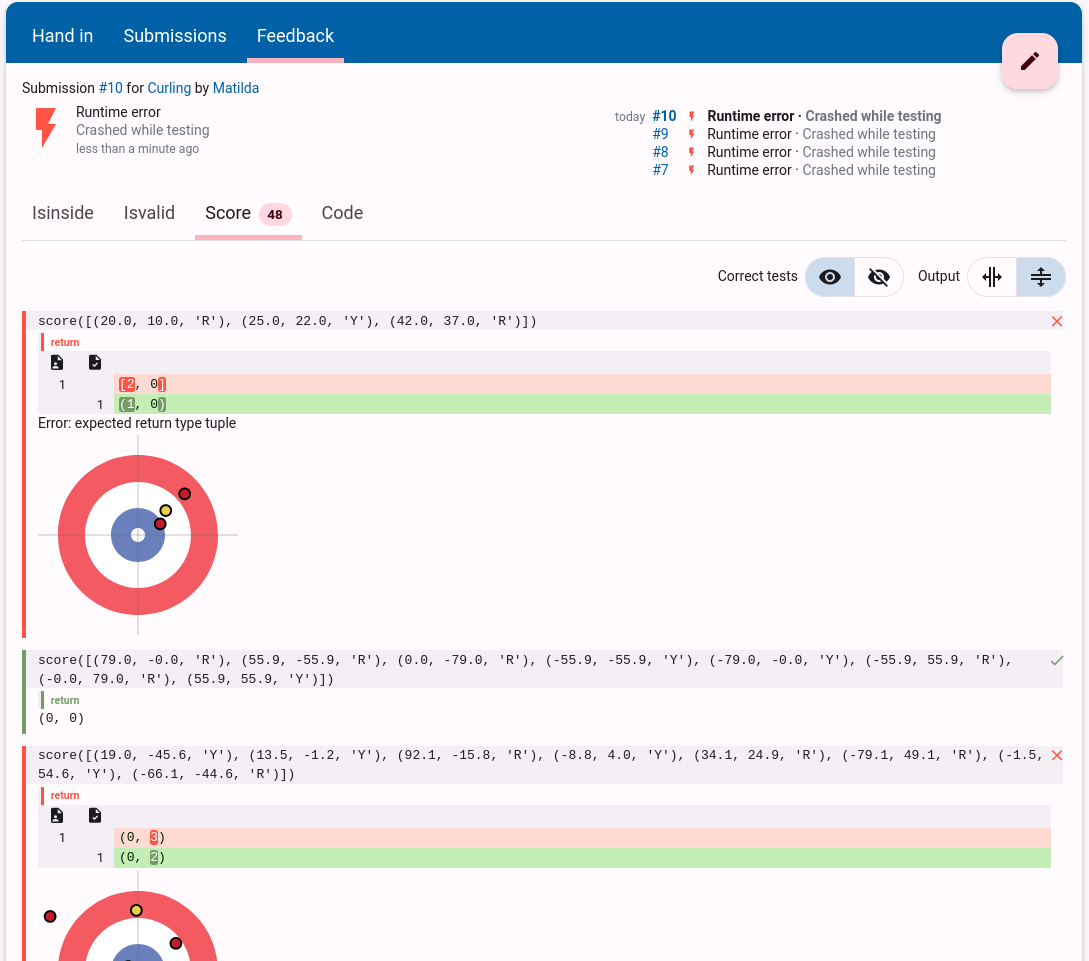}
  \caption{\Dodona{} rendering of feedback generated by the judge that assessed
    a submission of the Python programming assignment
    ``Curling''\protect\footnotemark. The judge split its feedback across three
    tabs, one for each function that needs to be implemented for this
    assignment: \texttt{isinside}, \texttt{isvalid} and \texttt{score}. All
    tests under the \texttt{isinside} and \texttt{isvalid} tabs passed, but 48
    tests under the \texttt{score} tab failed as can be seen immediately from
    the badge in the tab header. \Dodona{} also added a fourth tab ``Code'' that
    displays the source code of the submission with annotations added during
    automatic and/or manual assessment (Figure~\ref{fig:annotations}). Green/red
    vertical lines on the left reflect the grouping of test cases into execution
    contexts (here each execution context contains a single test
    case). \Dodona{} automatically highlighted the differences between the
    generated and expected return values of the first and third (failed) test
    case and the judge used unstructured HTML snippets to add a graphical
    representation (SVG) of the curling stone positions that are passed as
    arguments to the \texttt{score} function for these failed test cases. In
    addition to highlighting differences between the generated and expected
    return values of the first (failed) test case, the judge also added an
    unstructured text snippet that indicates that a \texttt{tuple} was expected
    (not a \texttt{list}).}\label{fig:feedback}
\end{figure*}

Where automatic assessment and feedback generation is outsourced to the judge
linked to an assignment, \Dodona{} itself takes up the responsibility for
rendering the feedback. This frees judge developers from putting effort in
feedback rendering and gives a coherent look-and-feel even for students that
solve programming assignments assessed by different judges. Because the way
feedback is presented is very important \citep{mani_better_2014}, we took great
care in designing how feedback is displayed to make its interpretation as easy
as possible (Figure~\ref{fig:feedback}). Differences between generated and
expected output are automatically highlighted for each failed test
\citep{myers_anond_1986}, and users can swap between displaying the output lines
side-by-side or interleaved to make differences more comparable. We even provide
specific support for highlighting differences between tabular data such as
CSV-files, database tables and dataframes. Users have the option to dynamically
hide contexts whose test cases all succeeded, allowing them to immediately
pinpoint reported mistakes in feedback that contains lots of succeeded test
cases. To ease debugging the source code of submissions for Python assignments,
the Python Tutor \citep{guo_online_2013} can be launched directly from any
context with a combination of the submitted source code and the test code from
the context. Students typically report this as one of the most useful features
of \Dodona{}.

\footnotetext{\Durl{https://dodona.ugent.be/en/activities/2051796188/}}

\subsection{Content management}

\begin{figure*}
  \centering
  \includegraphics[width=0.5\linewidth]{\Danon{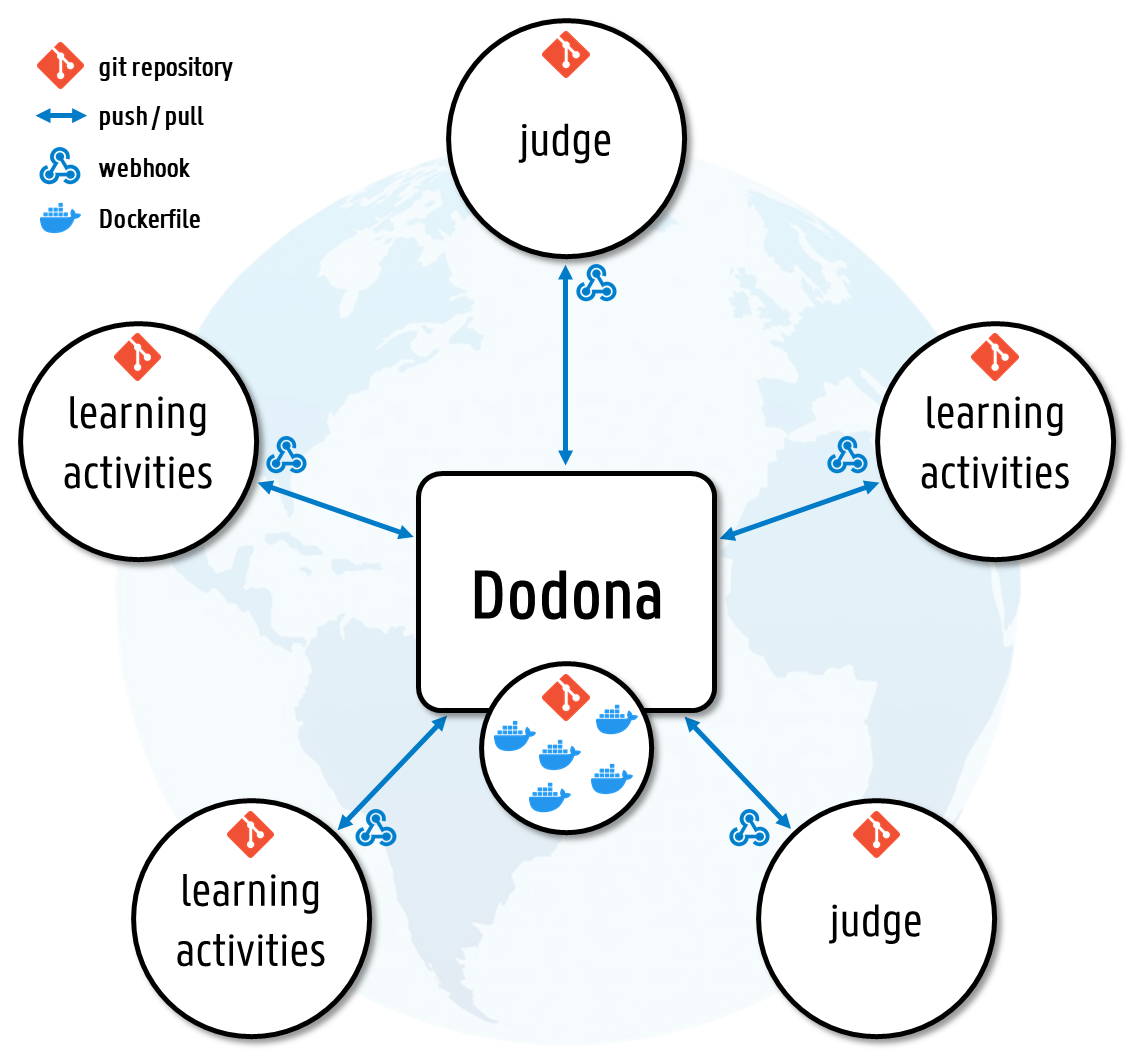}{repositories-anon.png}}
  \caption{Distributed content management model that allows to seamlessly
    integrate custom learning activities (reading activities and programming
    assignments with support for automated assessment) and judges (frameworks
    for automated assessment) into \Dodona{}. Content creators manage their content
    in external git repositories, keep ownership over their content, control who
    can co-create, and set up webhooks to automatically synchronize any changes
    with the content as published on \Dodona{}.}\label{fig:repositories}
\end{figure*}

Where courses are created and managed in \Dodona{} itself, other content is
managed in external git \textbf{repositories}
(Figure~\ref{fig:repositories}). In this distributed content management model, a
repository either contains a single judge or a collection of learning
activities: reading activities and/or programming
assignments\footnote{\Durl{https://docs.dodona.be/en/guides/teachers/new-exercise-repo/}}. Setting
up a \textbf{webhook} for the repository guarantees that any changes pushed to
its default branch are automatically and immediately synchronized with
\Dodona{}. This even works without the need to make repositories public, as they
may contain information that should not be disclosed such as programming
assignments that are under construction, contain model solutions, or will be
used during tests or exams. Instead, a \textbf{\Dodona{} service account} must
be granted push/pull access to the repository. Some settings of a learning
activity can be modified through the web interface of \Dodona{}, but any changes
are always pushed back to the repository in which the learning activity is
configured so that it always remains the master copy.

We experienced that working with git has a learning curve for some content
creators, who otherwise definitely benefit from its version control
capabilities. Due to the distributed nature of content management, creators also
keep ownership over their content and control who may co-create. After all,
access to a repository is completely independent from access to its learning
activities that are published in \Dodona{}. The latter is part of the
configuration of learning activities, with the option to either share learning
activities so that all teachers can include them in their courses or to restrict
inclusion of learning activities to courses that are explicitly granted
access. \Dodona{} automatically stores metadata about all learning activities such
as content type, natural language, programming language and repository to
increase their findability in our large collection. Learning activities may also
be tagged with additional \textbf{labels} as part of their configuration.

\DeclareFixedFootnote{\directoryStructure}{\Durl{https://docs.dodona.be/en/references/exercise-directory-structure/}}

Any repository containing learning activities must have a predefined directory
structure\directoryStructure{}. Directories
that contain a learning activity also have their own internal directory
structure\directoryStructure{}
that includes a \textbf{description} in Markdown or HTML\@. Descriptions may
reference data files and multimedia content included in the repository, and such
content can be shared across all learning activities in the repository. Embedded
images are automatically encapsulated in a responsive lightbox to improve
readability\footnote{\Durl{https://docs.dodona.be/en/references/exercise-description/}}. Mathematical
formulas in descriptions are supported through MathJax
\citep{cervone_mathjax_2012}.

While reading activities only consist of descriptions, programming assignments
need an additional \textbf{assessment configuration} that sets a programming
language and a
judge\footnote{\Durl{https://docs.dodona.be/en/references/exercise-config/}}. The
configuration may also set a Docker image, a time limit, a memory limit and
grant Internet access to the container that is instantiated from the image, but
these settings have proper default values. Judges, for example, have a default
image that is used if the configuration of a programming assignment does not
specify one explicitly. \Dodona{} builds the available images from Dockerfiles
specified in a separate git
repository\footnote{\Durl{https://github.com/dodona-edu/docker-images}}. The
configuration might also provide additional \textbf{assessment resources}: files
made accessible to the judge during
assessment\directoryStructure{}. The
specification of how these resources must be structured and how they are used
during assessment is completely up to the judge developers. Finally, the
configuration might also contain \textbf{boilerplate code}: a skeleton students
can use to start the implementation that is provided in the code editor along
with the description.

Taken together, a Docker image, a judge and a programming assignment
configuration (including both a description and an assessment configuration)
constitute a \textbf{task package} as defined by
\citet{verhoeff_programming_2008}: a unit \Dodona{} uses to render the
description of the assignment and to automatically assess its
submissions. However, \Dodona{}'s layered design embodies the separation of
concerns \citep{laplante_what_2007} needed to develop, update and maintain the
three modules in isolation and to maximize their reuse: multiple judges can use
the same Docker image and multiple programming assignments can use the same
judge. Related to this, an explicit design goal for judges is to make the
assessment configuration for individual assignments as lightweight as
possible. After all, minimal configurations reduce the time and effort teachers
and instructors need to create programming assignments that support automated
assessment. Sharing of data files and multimedia content among the programming
assignments in a repository also implements the inheritance mechanism for
\textbf{bundle packages} as hinted by \citep{verhoeff_programming_2008}. Another
form of inheritance is specifying default assessment configurations at the
directory level, which takes advantage of the hierarchical grouping of learning
activities in a repository to share common settings.

\subsection{Internationalization and localization}

\textbf{Internationalization} (i18n) is a shared responsibility between
\Dodona{}, learning activities and judges. All boilerplate text in the user
interface that comes from \Dodona{} itself is supported in English and Dutch,
and users can select their preferred language. Content creators can specify
descriptions of learning activities in both languages, and \Dodona{} will render
a learning activity in the user's preferred language if available. When users
submit solutions for a programming assignment, their preferred language is
passed as submission metadata to the judge. It's then up to the judge to take
this information into account while generating feedback.

\Dodona{} always displays \textbf{localized deadlines} based on a time zone setting in
the user profile, and users are warned when the current time zone detected by
their browser differs from the one in their profile.

\subsection{Questions, answers and code reviews}

\begin{figure*}
  \centering
  \includegraphics[width=\linewidth]{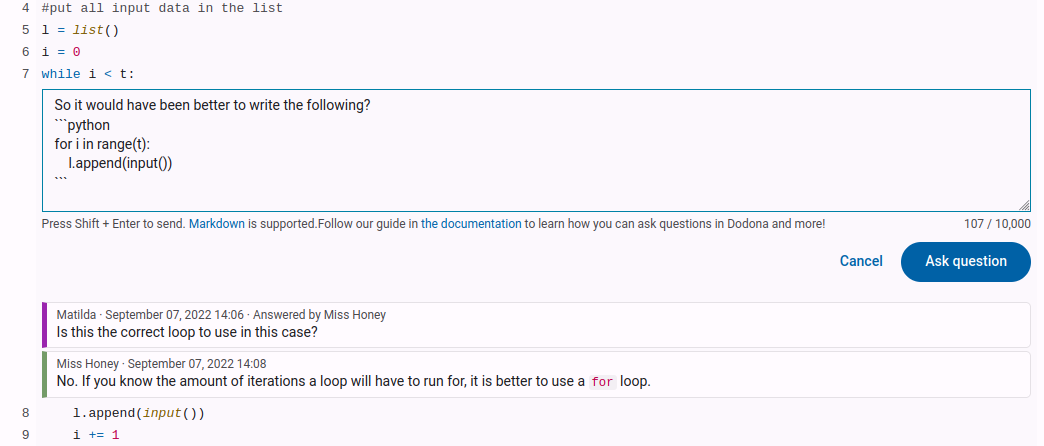}
  \caption{A student (Matilda) previously asked a question that has already
    been answered by her teacher (Miss Honey). Based on this response, the
    student is now asking a follow-up question that can be formatted using
    Markdown.}\label{fig:question}
\end{figure*}

A downside of using discussion forums in programming courses is that students
can ask questions about programming assignments that are either disconnected
from their current implementation or contain code snippets that may give away
(part of) the solution to other students
\citep{nandi_evaluating_2012}. \Dodona{} therefore allows students to address
teachers with questions they directly attach to their submitted source code. We
support both general questions and questions linked to specific lines of their
submission (Figure~\ref{fig:question}). Questions are written in Markdown (e.g.,
to include markup, tables, syntax highlighted code snippets or multimedia), with
support for MathJax (e.g., to include mathematical formulas).

\begin{figure*}
  \centering
  \includegraphics[width=\linewidth]{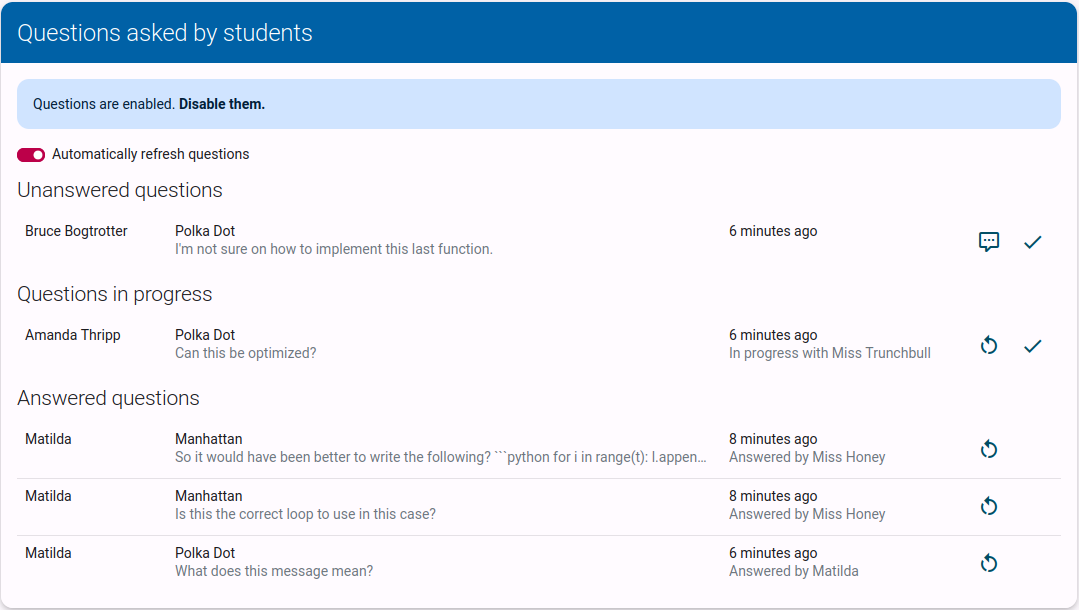}
  \caption{Live updated dashboard showing all incoming questions in a course
    while asking questions is enabled. Questions are grouped into three
    categories: unanswered, in progress and answered.}\label{fig:questions}
\end{figure*}

Teachers are notified whenever there are pending questions
(Figure~\ref{fig:course}). They can process these questions from a dedicated
dashboard with live updates (Figure~\ref{fig:questions}). The dashboard
immediately guides them from an incoming question to the location in the source
code of the submission it relates to, where they can answer the question in a
similar way as students ask questions. To avoid questions being inadvertently
handled simultaneously by multiple teachers, they have a three-state lifecycle:
pending, in progress and answered. In addition to teachers changing question
states while answering them, students can also mark their own questions as being
answered. The latter might reflect the rubber duck debugging
\citep{hunt_pragmatic_1999} effect that is triggered when students are forced to
explain a problem to someone else while asking questions in \Dodona{}. Teachers
can (temporarily) disable the option for students to ask questions in a course,
e.g.\ when a course is over or during hands-on sessions or exams when students
are expected to ask questions face-to-face rather than online.

Manual source code annotations from students (questions) and teachers (answers)
are rendered in the same way as source code annotations resulting from automated
assessment. They are mixed in the source code displayed in the ``Code'' tab,
showing their complementary nature. It is not required that students take the
initiative for the conversation. Teachers can also start adding source code
annotations while reviewing a submission. Such code reviews will be used as a
building block for manual assessment.

\subsection{Manual assessment}

Teachers can create an \textbf{evaluation} for a series to manually assess
student submissions for its programming assignments after a specific period,
typically following the deadline of some homework, an intermediate test or a
final exam. The evaluation embodies all programming assignments in the series
and a group of students that submitted solutions for these assignments. Because
a student may have submitted multiple solutions for the same assignment, the
last submission before a given deadline is automatically selected for each
student and each assignment in the evaluation. This automatic selection can be
manually overruled afterwards. The evaluation deadline defaults to the deadline
set for the associated series, if any, but an alternative deadline can be
selected as well.

Evaluations support \textbf{two-way navigation} through all selected
submissions: per assignment and per student. For evaluations with multiple
assignments, it is generally recommended to assess per assignment and not per
student, as students can build a reputation throughout an assessment
\citep{malouff_bias_2016}. As a result, they might be rated more favorably with
a moderate solution if they had excellent solutions for assignments that were
assessed previously, and vice versa \citep{malouff_risk_2013}. Assessment per
assignment breaks this reputation as it interferes less with the quality of
previously assessed assignments from the same student. Possible bias from the
same sequence effect is reduced during assessment per assignment as students are
visited in random order for each assignment in the evaluation. In addition,
\textbf{anonymous mode} can be activated as a measure to eliminate the actual or
perceived halo effect conveyed through seeing a student's name during assessment
\citep{lebuda_tell_2013}. While anonymous mode is active, all students are
automatically pseudonymized. Anonymous mode is not restricted to the context of
assessment and can be used across \Dodona{}, for example while giving in-class
demos.

\begin{figure*}
  \centering
  \includegraphics[width=\linewidth]{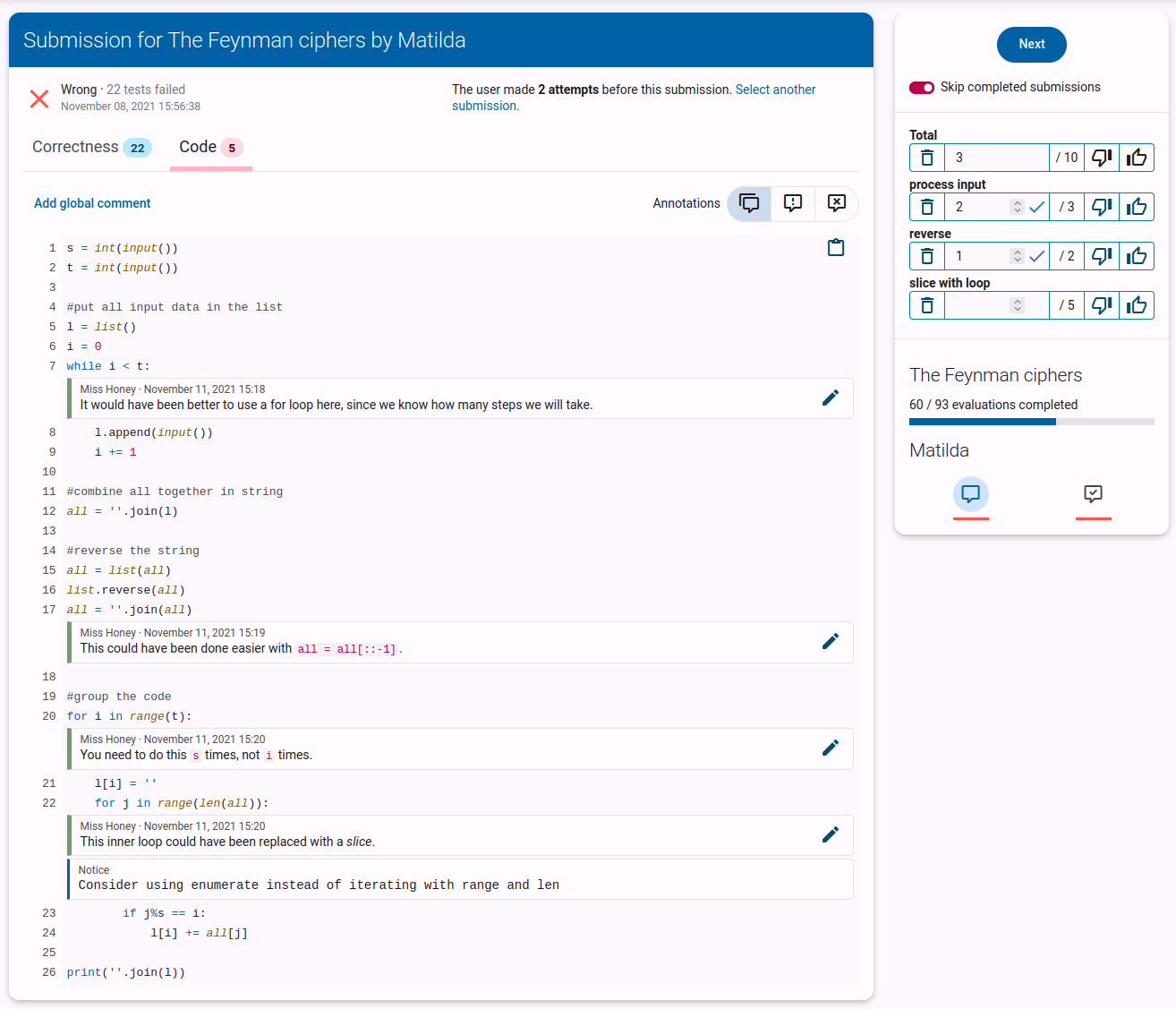}
  \caption{Manual assessment of a submission: a teacher (Miss Honey) is giving
    feedback on the source code by adding inline annotations and is grading the
    submission by filling up the scoring rubric that was set up for the
    programming assignment ``The Feynman ciphers''.}\label{fig:annotations}
\end{figure*}

When reviewing a selected submission from a student, assessors have direct
access to the feedback that was previously generated during automated
assessment: source code annotations in the ``Code'' tab and other structured and
unstructured feedback in the remaining tabs. Moreover, next to the feedback that
was made available to the student, the specification of the assignment may also
add feedback generated by the judge that is only visible to the
assessor. Assessors might then complement the assessment made by the judge by
adding \textbf{source code annotations} as formative feedback and by
\textbf{grading} the evaluative criteria in a scoring rubric as summative
feedback (Figure~\ref{fig:annotations}). Previous annotations can be reused to
speed up the code review process, because remarks or suggestions tend to recur
frequently when reviewing submissions for the same assignment. Grading requires
setting up a specific \textbf{scoring rubric} for each assignment in the
evaluation, as a guidance for evaluating the quality of submissions
\citep{dawson_assessment_2017,popham_whats_1997}. The evaluation tracks which
submissions have been manually assessed, so that analytics about the assessment
progress can be displayed and to allow multiple assessors working simultaneously
on the same evaluation, for example one (part of a) programming assignment each.

An important difference with automated assessment is that the feedback from
manual assessment is \textbf{delivered asynchronously}. Students can only see
manual source code annotations and grades after a teacher has explicitly
published the feedback from an evaluation. This allows publishing the feedback
from manual assessment to all students at once. In response, students can start
asking questions about the feedback and grades they received, given that the
option to ask questions is enabled in the course. An evaluation also provides
teachers with \textbf{summary statistics and overview reports} that can be
exported to external grade books.

\subsection{Implementation}

\Dodona{} has a multi-tier service architecture, which delegates separate parts of
the application to dedicated servers or virtual machines. This increases
robustness to failure and improves reliability and scalability when serving
hundreds of concurrent users. More specifically, the web server, database
(MySQL), caching system (Memcached) and Python Tutor each run on their own
machine. In addition, a scalable pool of interchangeable worker servers are
available to automatically assess incoming student submissions.

The web server is the only public-facing part of \Dodona{}, running a Ruby on Rails
web application that is available on GitHub under the permissive MIT open-source
license\footnote{\Durl{https://github.com/dodona-edu}}. The user interface is
built using Bootstrap\footnote{\url{https://getbootstrap.com/}} and follows the
Google Material Design specifications\footnote{\url{https://material.io/}}. This
ensures a coherent and accessible design that works in all modern web
browsers. Dark mode is supported for programmers who fear that light attracts
bugs. Next to the graphical web interface, \Dodona{} also provides an
application programming interface (API) using JSON, allowing external
applications and scripts to interoperate with \Dodona{}.

Software developers outside the core \Dodona{} development team have used the
API to implement IDE plug-ins for Visual Studio
Code\footnote{\Durl{https://docs.dodona.be/en/guides/vs-code-extension/}},
DrRacket\footnote{\Durl{http://soft.vub.ac.be/SCPI/}} and JetBrains IDEs like
IntelliJ IDEA, PyCharm or
Webstorm\footnote{\Durl{https://docs.dodona.be/en/guides/pycharm-plugin/}}. These
IDE extensions directly embed support for submitting solutions and feedback
notifications into the environment where students work on their assignments. The
plug-ins hook onto the \Dodona{} API with secure authentication through API
tokens. In addition, \Dodona{} provides LTI
1.3\footnote{\url{http://www.imsglobal.org/activity/learning-tools-interoperability}}
support for seamless integration of courses and learning activities in external
learning management systems through single sign-on and deep linking.

\subsection{Reliability and security}

\Dodona{} needs to operate in a challenging environment where students
simultaneously submit untrusted code to be executed on its servers (``remote
code exection by design'') and expect automatically generated feedback, ideally
within a few seconds. Many design decisions are therefore aimed at maintaining
and improving the reliability and security of its systems.

With respect to reliability, \Dodona{} minimizes the risk that its systems get
overwhelmed by requests and submissions. This is especially important during
tests and exams when large groups of students simultaneously submit solutions
during the final minutes before a deadline. System overload is mitigated through
a multi-tier architecture that puts student submissions in a job queue upon
arrival. Worker servers then assess submissions on a first come, first serve
basis. A sudden influx of submissions may cause the job queue to become longer
temporarily, but apart from slightly longer waiting times for feedback delivery,
this has no other adverse side-effects. To prevent resource hogging of
individual submissions, all source code is assessed server-side in separate
Docker containers with strict limits on disk, memory, and CPU usage. These
limits restrict the impact of processing individual submissions on other jobs
running on the same server. If, despite everything, things do go wrong and a
worker server succumbs under the load, other worker servers from the pool will
pick up the slack at the cost of only slightly longer wait times. Moreover, in
such cases, strict operations monitoring will notify server administrators at
once.

Security is a much broader topic and significantly harder to fully
safeguard. Looking at the OWASP Top Ten web application security
risks\footnote{\url{https://owasp.org/www-project-top-ten/}}, almost all of them
apply to \Dodona{} and mandate a mitigation strategy. We prevent broken access
control by employing the widely-used
pundit\footnote{\url{https://github.com/varvet/pundit}} library to make account
permissions clearly readable in policy files. Access control is checked upon
each request. Standard implementations of security algorithms and standard
configurations of applications are used to preclude cryptographic
failures. ActiveRecord and standard CSRF protection from Rails avert injection
attacks. The additional risk of injections through user-provided content is
mitigated by sandboxing content in iframes and using a strict content security
policy (CSP). Content that can not be sandboxed properly is sanitized using
standard Rails support. We follow standard Rails development patterns to ensure
a secure design for \Dodona{}. All software updates are reviewed by at least two
other developers familiar with the \Dodona{} codebase and tested automatically
through continuous integration pipelines. The codebase is screened by static
source code analysis to detect known
vulnerabilities. Dependabot\footnote{\url{https://github.com/dependabot}}
automatically monitors external dependencies to assure that no outdated
libraries or libraries with known vulnerabilities are used
\citep{alfadel_use_2021}. This also applies to the Docker images in which
student submissions are assessed. Identification and authentication failures are
bypassed through decentralized authentication and authorization. Finally, we use
Grafana\footnote{\url{https://grafana.com/}} to actively monitor the application
and the servers through internal dashboards that provide up-to-date status
reports and to automatically trigger alerts if system indicators show abnormal
trends or events. Additional alerts are triggered each time an error occurs in
the \Dodona{} web application, informing us about bugs or other problems.

\section{Managing active learning in introductory programming: a case study}\label{managing}

Since the academic year 2011--2012 we have organized an introductory Python
course at \Danon{Ghent University (Belgium)}{our university} with a strong focus on active and online
learning. Initially the course was offered twice a year in the first and second
term, but from academic year 2014--2015 onwards it was only offered in the first
term. The course is taken by a mix of undergraduate, graduate, and postgraduate
students enrolled in various study programmes (mainly formal and natural
sciences, but not computer science), with 442 students enrolled for the
2021--2022 edition\footnote{\Durl{https://dodona.ugent.be/en/courses/773/}}.

\subsection{Course structure}

\begin{figure*}
  \centering
  \includegraphics[width=0.8\linewidth]{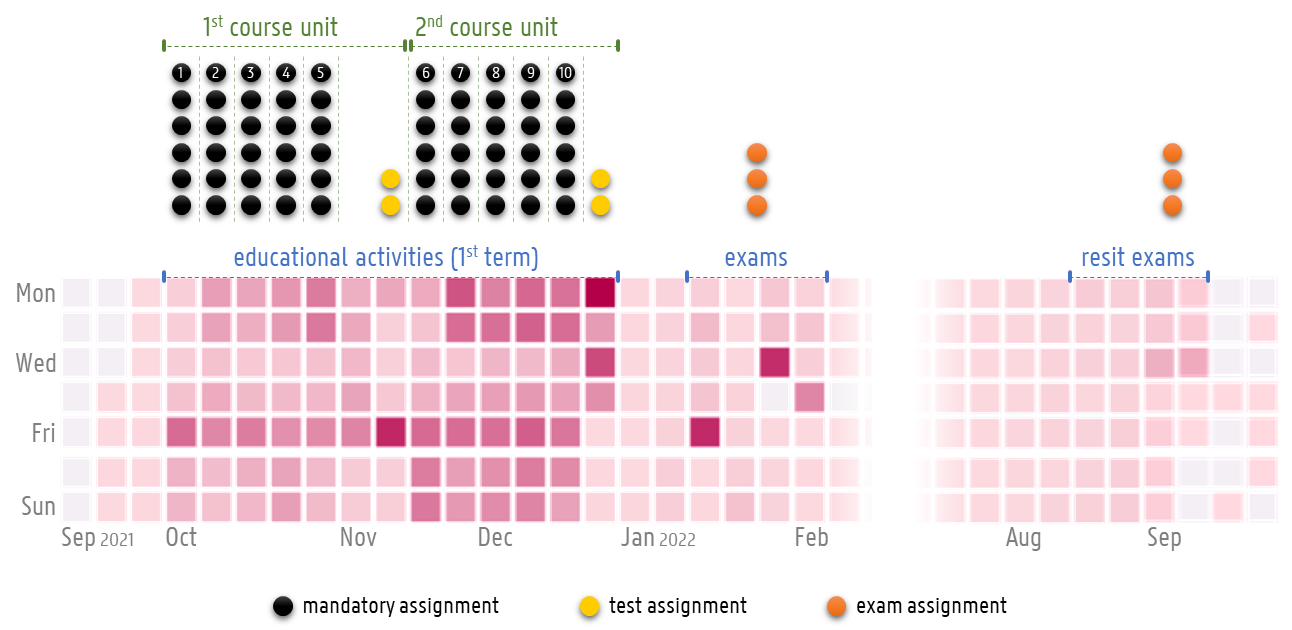}
  \caption{\textbf{Top}: Structure of the Python course that runs each
    academic year across a 13-week term (September-December). Programming
    assignments from the same \Dodona{} series are stacked vertically. Students
    submit solutions for ten series with six mandatory assignments, two tests
    with two assignments and an exam with three assignments. They can also take
    a resit exam with three assignments in August/September if they failed the
    first exam in January. Each series of mandatory assignments has a dedicated
    topic: (1) variables, expressions and statements, (2) conditional
    statements, (3) loops, (4) strings, (5) functions, (6) lists and tuples, (7)
    more about functions and modules, (8) sets and dictionaries, (9) text files,
    (10) object-oriented programming. \textbf{Bottom}: Heatmap from \Dodona{}
    learning analytics page showing distribution per day of all 331~734
    solutions submitted during the 2021--2022 edition of the course (442
    students). The darker the color, the more solutions were submitted that
    day. A lighter shade of fuchsia means few solutions were submitted that
    day. A light gray square means no solutions were submitted that day. Weekly
    lab sessions for different groups on Monday afternoon, Friday morning and
    Friday afternoon, where we can see darker squares. Weekly deadlines for
    mandatory assignments on Tuesdays at 22:00. Three exam sessions for
    different groups in January. Low activity in exam periods, except for days
    where an exam was taken. The course is not taught in the second term, so
    this low-activity period was collapsed. Two more exam sessions for different
    groups in August/September, granting an extra chance to students who failed
    on their exam in January.}\label{fig:course-structure}
\end{figure*}

Each course edition has a fixed structure, with 13 weeks of educational
activities subdivided in two successive instructional units that each cover five
topics of the Python programming language --- one topic per week --- followed by
a graded test about all topics covered in the unit
(Figure~\ref{fig:course-structure}). The final exam at the end of the term
evaluates all topics covered in the entire course. Students who fail the course
during the first exam in January can take a resit exam in August/September that
gives them a second chance to pass the exam.

Each week in which a new programming topic is covered, students must try to
solve six programming assignments on that topic before a deadline one week
later. That results in 60 mandatory assignments across the semester. Following
the flipped classroom strategy \citep{akcayir_flipped_2018,bishop_flipped_2013},
students prepare themselves to achieve this goal by reading the textbook
chapters covering the topic. Lectures are interactive programming sessions that
aim at bridging the initial gap between theory and practice, advancing concepts,
and engaging in collaborative learning \citep{tucker_flipped_2012}. Along the
same lines, the first assignment for each topic is an ISBN-themed programming
challenge whose model solution is shared with the students, together with an
instructional video that works step-by-step towards the model solution. As soon
as students feel they have enough understanding of the topic, they can start
working on the five remaining mandatory assignments. Students can work on their
programming assignments during weekly computer labs, where they can collaborate
in small groups and ask help from teaching assistants. They can also work on
their assignments and submit solutions outside lab sessions. In addition to the
mandatory assignments, students can further elaborate on their programming
skills by tackling additional programming exercises they select from a pool of
over 850 exercises linked to the ten programming topics. Submissions for these
additional exercises are not taken into account in the final grade.

\subsection{Assessment, feedback and grading}

We use the online learning environment \Dodona{} to promote active learning
through problem solving \citep{prince_does_2004}. Each course edition has its
own dedicated course in \Dodona{}, with a learning path containing all
mandatory, test and exam assignments, grouped into series with corresponding
deadlines. Mandatory assignments for the first unit are published at the start
of the semester, and those for the second unit after the test of the first
unit. For each test and exam we organize multiple sessions for different groups
of students. Assignments for test and exam sessions are provided in a hidden
series that is only accessible for students participating in the session using a
shared secret link. The test and exam assignments are published afterwards for
all students, when grades are announced. Students can see class progress when
working on their mandatory assignments to nudge them to avoid
procrastination. Only teachers can see class progress for test and exam series
so as not to accidentally stress out students. For the same reason, we
intentionally organize tests and exams following exactly the same procedure, so
that students can take high-stake exams in a familiar context and adjust their
approach based on previous experiences. The only difference is that test
assignments are not as hard as exam assignments, as students are still in the
midst of learning programming skills when tests are taken.

Students are stimulated to use an integrated development environment (IDE) to
work on their programming assignments. IDEs bundle a battery of programming
tools to support today's generation of software developers in writing, building,
running, testing and debugging software. Working with such tools can be a true
blessing for both seasoned and novice programmers, but there is no silver bullet
\citep{brooks_no_1987}. Learning to code remains inherently hard
\citep{kelleher_alice2_2002} and consists of challenges that are different to
reading and learning natural languages \citep{fincher_what_1999}. As an
additional aid, students can continuously submit (intermediate) solutions for
their programming assignments and immediately receive automatically generated
feedback upon each submission, even during tests and exams. Guided by that
feedback, they can track potential errors in their code, remedy them and submit
updated solutions. There is no restriction on the number of solutions that can
be submitted per assignment. All submitted solutions are stored, but for each
assignment only the last submission before the deadline is taken into account to
grade students. This allows students to update their solutions after the
deadline (i.e.\ after model solutions are published) without impacting their
grades, as a way to further practice their programming skills. One effect of
active learning, triggered by mandatory assignments with weekly deadlines and
intermediate tests, is that most learning happens during the term
(Figure~\ref{fig:course-structure}). In contrast to other courses, students do
not spend a lot of time practicing their coding skills for this course in the
days before an exam. We want to explicitly motivate this behavior, because we
strongly believe that one cannot learn to code in a few days' time
\citep{peter_norvig_teach_2001}.

We originally developed a custom Python judge for SPOJ (Sphere Online Judge) to
automatically assess student submissions for our own collection of programming
assignments \citep{kosowski_application_2008}. However, after five years of
running the course, we felt the shortcomings of SPOJ or other existing online
learning environments prevented us from expanding our vision on active learning
and providing rich feedback just-in-time and in a scalable way. In response, we
started developing \Dodona{} in spring 2016, ported our Python judge and
collection of programming assignments over the summer, and ran our first Python
course with \Dodona{} during the first term of academic year 2016--2017. Until this
day, structurally designing and modeling the Python course and its pedagogy
remains a driving force for further extending \Dodona{} and for validating novel
or improved features in educational practice. Along the way, other computer
science, data science and statistics courses adopted \Dodona{} with a variation of
learning contexts, programming languages and assessment requirements. But more
on this in the next section.

For the assessment of tests and exams, we follow the line of thought that human
expert feedback through source code annotations is a valuable complement to
feedback coming from automated assessment, and that human interpretation is an
absolute necessity when it comes to grading
\citep{ala-mutka_survey_2005,jackson_grading_1997,staubitz_towards_2015}. We
shifted from paper-based to digital code reviews and grading when support for
manual assessment was released in version 3.7 of \Dodona{} (summer
2020). Although online reviewing positively impacted our productivity, the
biggest gain did not come from an immediate speed-up in the process of
generating feedback and grades compared to the paper-based approach. While
time-on-task remained about the same, our online source code reviews were much
more elaborate than what we produced before on printed copies of student
submissions. This was triggered by improved reusability of digital annotations
and the foresight of streamlined feedback delivery. Where delivering custom
feedback only requires a single click after the assessment of an evaluation has
been completed in \Dodona{}, it took us much more effort before to distribute
our paper-based feedback. Students were direct beneficiaries from more and
richer feedback, as observed from the fact that 75\% of our students looked at
their personalized feedback within 24 hours after it had been released, before
we even published grades in \Dodona{}. What did not change is the fact that we
complement personalized feedback with collective feedback sessions in which we
discuss model solutions for test and exam assignments, and the low numbers of
questions we received from students on their personalized feedback. As a future
development, we hope to reduce the time spent on manual assessment through
improved computer-assisted reuse of digital source code annotations in \Dodona{}.

We accept to primarily rely on automated assessment as a first step in providing
formative feedback while students work on their mandatory assignments. After
all, a back-of-the-envelope calculation tells it would take us 72 full-time
equivalents (FTE) to generate equivalent amounts of manual feedback for
mandatory assignments compared to what we do for tests and exams. In addition to
volume, automated assessment also yields the responsiveness needed to establish
an interactive feedback loop throughout the iterative software development
process while it still matters to students and in time for them to pay attention
to further learning or receive further assistance
\citep{gibbs_conditions_2005}. Automated assessment thus allows us to motivate
students working through enough programming assignments and to stimulate their
self-monitoring and self-regulated learning
\citep{pintrich_understanding_1995,schunk_self-regulation_1994}. It results in
triggering additional questions from students that we manage to respond to with
one-to-one personalized human tutoring, either synchronously during hands-on
sessions or asynchronously through \Dodona{}'s Q\&A module. We observe that
individual students seem to have a strong bias towards either asking for
face-to-face help during hands-on sessions or asking questions online. This
could be influenced by the time when they mainly work on their assignments, by
their way of collaboration on assignments, or by reservations because of
perceived threats to self-esteem or social embarrassment
\citep{karabenick_relationship_1991,newman_students_1993}.

In computing a final score for the course, we try to find an appropriate balance
between stimulating students to find solutions for programming assignments
themselves and collaborating with and learning from peers, instructors and
teachers while working on assignments. The final score is computed as the sum of
a score obtained for the exam (80\%) and a score for each unit that combines the
student's performance on the mandatory and test assignments (10\% per unit). We
use \Dodona{}'s grading module to determine scores for tests and exams based on
correctness, programming style, choice made between the use of different
programming techniques and the overall quality of the implementation. The score
for a unit is calculated as the score $s$ for the two test assignments
multiplied by the fraction $f$ of mandatory assignments the student has solved
correctly. A solution for a mandatory assignment is considered correct if it
passes all unit tests. Evaluating mandatory assignments therefore doesn't
require any human intervention, except for writing unit tests when designing the
assignments, and is performed entirely by our Python judge. In our experience,
most students traditionally perform much better on mandatory assignments
compared to test and exam assignments \citep{glass_fewer_2022}, given the
possibilities for collaboration on mandatory assignments.

\subsection{Open and collaborative learning environment}

We strongly believe that effective collaboration among small groups of students
is beneficial for learning \citep{prince_does_2004}, and encourage students to
collaborate and ask questions to tutors and other students during and outside
lab sessions. We also demonstrate how they can embrace collaborative coding and
pair programming services provided by modern integrated development environments
\citep{hanks_pair_2011,williams_support_2002}. But we recommend them to
collaborate in groups of no more than three students, and to exchange and
discuss ideas and strategies for solving assignments rather than sharing literal
code with each other. After all, our main reason for working with mandatory
assignments is to give students sufficient opportunity to learn topic-oriented
programming skills by applying them in practice and shared solutions spoil the
learning experience. The factor $f$ in the score for a unit encourages students
to keep finetuning their solutions for programming assignments until all test
cases succeed before the deadline passes. But maximizing that factor without
proper learning of programming skills will likely yield a low test score $s$ and
thus an overall low score for the unit, even if many mandatory exercises were
solved correctly.

\begin{figure*}
  \centering
  \includegraphics[width=\linewidth]{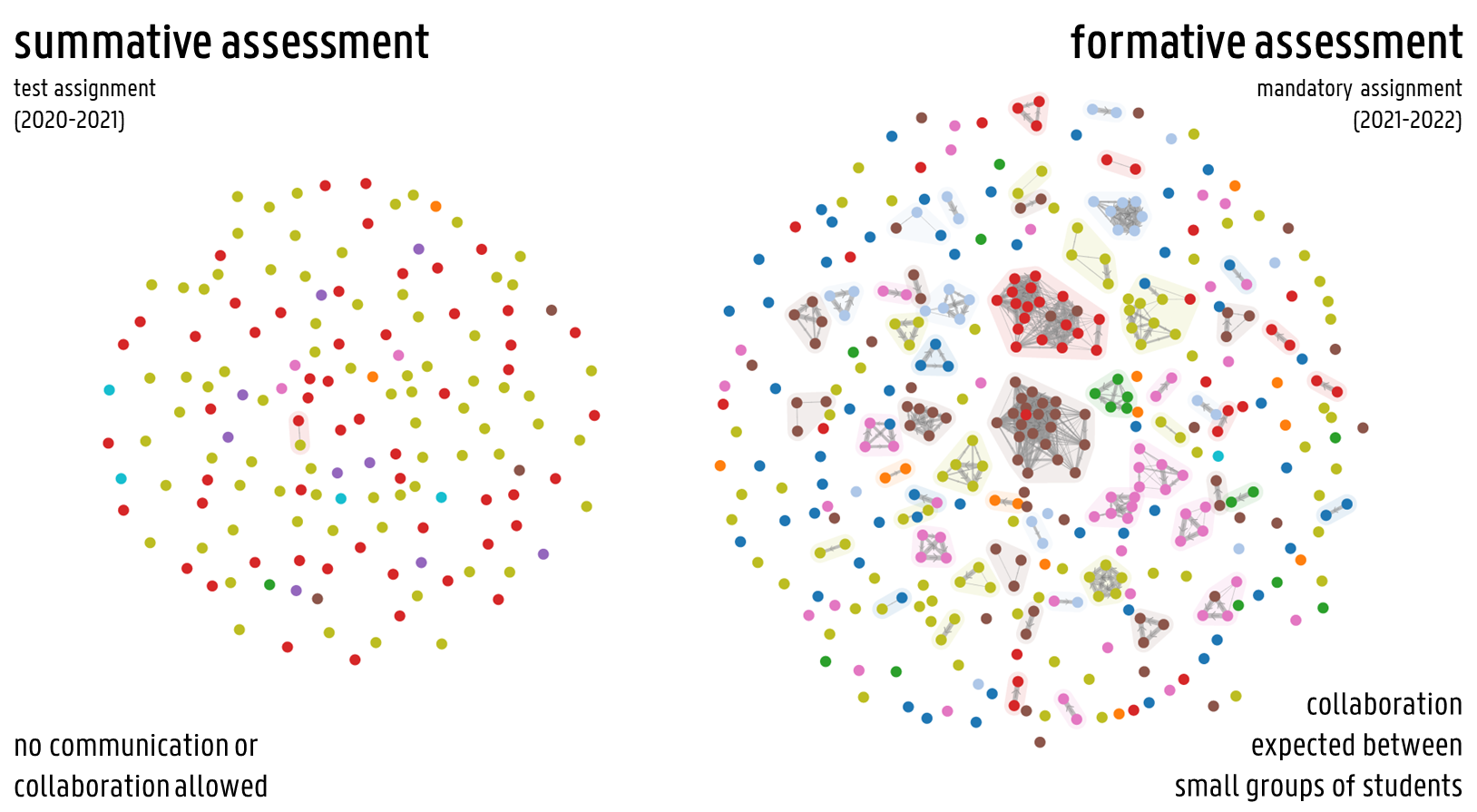}
  \caption{Dolos plagiarism graphs for the Python programming assignment
    ``$\pi$-ramidal constants'' that was created and used for a test of the
    2020--2021 edition of the course (left) and reused as a mandatory assignment
    in the 2021--2022 edition (right). Graphs constructed from the last
    submission before the deadline of 142 and 382 students respectively. Nodes
    represent student submissions and their colors represent study programmes as
    taken from user labels in \Dodona{}. Edges connect highly similar pairs of
    submissions, with similarity threshold set to 0.8 in both graphs. Edge
    directions are based on submission timestamps in \Dodona{}. Clusters of
    connected nodes are highlighted with a distinct background color and have
    one node with a solid border that indicates the first correct submission
    among all submissions in that cluster. All students submitted unique
    solutions during the test, except for two students who confessed they
    exchanged a solution during the test. Submissions for the mandatory
    assignment show that most students work either individually or in groups of
    two or three students, but we also observe some clusters of four or more
    students that exchanged solutions and submitted them with hardly any varying
    types and amounts of modifications. This case was used to warn students
    about the negative learning effect of copying solutions from each
    other.}\label{fig:plagiarism}
\end{figure*}

Fostering an open collaboration environment to work on mandatory assignments
with strict deadlines and taking them into account for computing the final score
is a potential promoter for plagiarism, but using it as a weight factor for the
test score rather than as an independent score item should promote learning by
avoiding that plagiarism is rewarded. It takes some effort to properly explain
this to students. We initially used Moss \citep{schleimer_winnowing_2003} and
now use Dolos \citep{maertens_dolos_2022} to monitor submitted solutions for
mandatory assignments, both before and at the deadline. The solution space for
the first few mandatory assignments is too small for linking high similarity to
plagiarism: submitted solutions only contain a few lines of code and the
diversity of implementation strategies is small. But at some point, as the
solution space broadens, we start to see highly similar solutions that are
reliable signals of code exchange among larger groups of students. Strikingly
this usually happens among students enrolled in the same study programme
(Figure~\ref{fig:plagiarism}). As soon as this happens --- typically in week 3
or 4 of the course --- plagiarism is discussed during the next lecture. Usually
this is a lecture about working with the string data type, so we can introduce
plagiarism detection as a possible application of string processing.

In an intermezzo entitled ``copy-paste $\neq$ learn to code'' we show students some
pseudonymized Dolos plagiarism graphs that act as mirrors to make them reflect
upon which node in the graph they could be (Figure~\ref{fig:plagiarism}). We stress that the
learning effect dramatically drops in groups of four or more students. Typically
we notice that in such a group only one or a few students make the effort to
learn to code, while the other students usually piggyback by copy-pasting
solutions. We make students aware that understanding someone else's code for
programming assignments is a lot easier than trying to find solutions
themselves. Over the years, we have experienced that a lot of students are
caught in the trap of genuinely believing that being able to understand code is
the same as being able to write code that solves a problem until they take a
test at the end of a unit. That's where the $s$ factor of the test score comes
into play. After all, the goal of summative tests is to evaluate if individual
students have acquired the skills to solve programming challenges on their own.

When talking to students about plagiarism, we also point out that the plagiarism
graphs are directed graphs, indicating which student is the potential source of
exchanging a solution among a cluster of students. We specifically address these
students by pointing out that they are probably good at programming and might
want to exchange their solutions with other students in a way to help their
peers. But instead of really helping them out, they actually take away learning
opportunities from their fellow students by giving away the solution as a
spoiler. Stated differently, they help maximize the factor $f$ but effectively
also reduce the $s$ factor of the test score, where both factors need to be high
to yield a high score for the unit. After this lecture, we usually notice a
stark decline in the amount of plagiarized solutions.

The goal of plagiarism detection at this stage is prevention rather than
penalisation, because we want students to take responsibility over their
learning. The combination of realizing that teachers and instructors can easily
detect plagiarism and an upcoming test that evaluates if students can solve
programming challenges on their own, usually has an immediate and persistent
effect on reducing cluster sizes in the plagiarism graphs to at most three
students. At the same time, the signal is given that plagiarism detection is one
of the tools we have to detect fraud during tests and exams. The entire group of
students is only addressed once about plagiarism, without going into detail
about how plagiarism detection itself works, because we believe that
overemphasizing this topic is not very effective and explaining how it works
might drive students towards spending time thinking on how they could bypass the
detection process --- time better spent on learning to code. Every
three or four years we see a persistent cluster of students exchanging code for
mandatory assignments over multiple weeks. If this is the case, we individually
address these students to point them again on their responsibilities, again
differentiating between students that share their solution and students that
receive solutions from others.

Tests and exams, on the other hand, are taken on-campus under human surveillance
and allow no communication with fellow students or other persons. Students can
work on their personal computers and get exactly two hours to solve two
programming assignments during a test, and three hours and thirty minutes to
solve three programming assignments during an exam. Tests and exams are ``open
book/open Internet'', so any hard copy and digital resources can be consulted
while solving test or exam assignments. Students are instructed that they can
only be passive users of the Internet: all information available on the Internet
at the start of a test or exam can be consulted, but no new information can be
added. When taking over code fragments from the Internet, students have to add a
proper citation as a comment in their submitted source code. After each test and
exam, we again use Moss/Dolos to detect and inspect highly similar code snippets
among submitted solutions and to find convincing evidence they result from
exchange of code or other forms of interpersonal communication
(Figure~\ref{fig:plagiarism}). If we catalog cases as plagiarism beyond
reasonable doubt, the examination board is informed to take further
action\Danon{ \citep{maertens_dolos_2022}}{}.

\subsection{Workload for running a course edition}

To organize ``open book/open Internet'' tests and exams that are valid and
reliable, we always create new assignments and avoid assignments whose solutions
or parts thereof are readily available online. At the start of a test or exam,
we share a secret link that gives students access to the assignments in a hidden
series on \Dodona{}.

For each edition of the course, mandatory assignments were initially a
combination of selected test and exam exercises reused from the previous edition
of the course and newly designed exercises. The former to give students an idea
about the level of exercises they can expect during tests and exams, and the
latter to avoid solution slippage. As feedback for the students we publish
sample solutions for all mandatory exercises after the weekly deadline has
passed. This also indicates that students must strictly adhere to deadlines,
because sample solutions are available afterwards. As deadlines are very clear
and adjusted to timezone settings in \Dodona{}, we never experience discussions
with students about deadlines.

After nine editions of the course, we felt we had a large enough portfolio of
exercises to start reusing mandatory exercises from four or more years ago
instead of designing new exercises for each edition. However, we still continue
to design new exercises for each test and exam. After each test and exam,
exercises are published and students receive manual reviews on the code they
submitted, on top of the automated feedback they already got during the test or
exam. But in contrast to mandatory exercises we do not publish sample solutions
for test and exam exercises, so that these exercises can be reused during the
next edition of the course. When students ask for sample solutions of test or
exam exercises, we explain that we want to give the next generation of students
the same learning opportunities they had.

So far, we have created more than 850 programming assignments for this
introductory Python course alone. All these assignments are publicly shared on
\Dodona{} as open educational resources
\citep{caswell_open_2008,downes_models_2007,hylen_open_2021,tuomi_open_2013,wiley_open_2014}. They
are used in many other courses on \Dodona{} (on average 10.8 courses per
assignment) and by many students (on average 503.7 students and 4801.5 submitted
solutions per assignment). We estimate that it takes about 10 person-hours on
average to create a new assignment for a test or an exam: 2 hours for ideation,
30 minutes for implementing and tweaking a sample solution that meets the
educational goals of the assignment and can be used to generate a test suite for
automated assessment, 4 hours for describing the assignment (including
background research), 30 minutes for translating the description from Dutch into
English, one hour to configure support for automated assessment, and another 2
hours for reviewing the result by some extra pair of eyes.

Generating a test suite usually takes 30 to 60 minutes for assignments that can
rely on basic test and feedback generation features that are built into the
judge. The configuration for automated assessment might take 2 to 3 hours for
assignments that require more elaborate test generation or that need to extend
the judge with custom components for dedicated forms of assessment
(e.g.\ assessing non-deterministic behavior) or feedback generation
(e.g.\ generating visual feedback). \citet{keuning_systematic_2018} found that
publications rarely describe how difficult and time-consuming it is to add
assignments to automated assessment platforms, or even if this is possible at
all. The ease of extending \Dodona{} with new programming assignments is
reflected by more than 10 thousand assignments that have been added to the platform so
far. Our experience is that configuring support for automated assessment only
takes a fraction of the total time for designing and implementing assignments
for our programming course, and in absolute numbers stays far away from the one
person-week reported for adding assignments to Bridge
\citep{bonar_bridge_1988}. Because the automated assessment infrastructure of
\Dodona{} provides common resources and functionality through a Docker container
and a judge, the assignment-specific configuration usually remains
lightweight. Only around 5\% of the assignments need extensions on top of the
built-in test and feedback generation features of the judge.

So how much effort does it cost us to run one edition of our programming course?
For the most recent 2021--2022 edition we estimate about 34 person-weeks in
total (Table~\ref{tab:workload}), the bulk of which is spent on on-campus
tutoring of students during hands-on sessions (30\%), manual assessment and
grading (22\%), and creating new assignments (21\%). About half of the workload
(53\%) is devoted to summative feedback through tests and exams: creating
assignments, supervision, manual assessment and grading. Most of the other work
(42\%) goes into providing formative feedback through on-campus and online
assistance while students work on their mandatory assignments. Out of 2215
questions that students asked through \Dodona{}'s online Q\&A module, 1983
(90\%) were answered by teaching assistants and 232 (10\%) were marked as
answered by the student who originally asked the question. Because automated
assessment provides first-line support, the need for human tutoring is already
heavily reduced. We have drastically cut the time we initially spent on
mandatory assignments by reusing existing assignments and because the Python
judge is stable enough to require hardly any maintenance or further development.

\begin{table}
  \centering
  \begin{tabular}{lr}
    \toprule
    \textbf{Task} & \textbf{Estimated workload (person-hours)} \\
    \midrule
    \textbf{Instruction: lectures} & \textbf{60} \\
    \midrule
    \textbf{Formative assessment: mandatory assignments} & \textbf{540} \\
    \hspace{1em} Select assignments & 10 \\
    \hspace{1em} Review selected assignments & 30 \\
    \hspace{1em} Tips \& tricks & 10 \\
    \hspace{1em} Automated assessment & 0 \\
    \hspace{1em} On-campus tutoring: hands-on sessions & 390 \\
    \hspace{1em} Online tutoring: answering questions in Q\&A module & 100 \\
    \midrule
    \textbf{Summative assessment: tests \& exams} & \textbf{690} \\
    \hspace{1em} Create new assignments for tests and exams & 270 \\
    \hspace{1em} Supervise tests and exams & 130 \\
    \hspace{1em} Automated assessment & 0 \\
    \hspace{1em} Manual assessment: code reviewing and grading & 288 \\
    \hspace{1em} Plagiarism detection & 2 \\
    \midrule
    \textbf{Total} & \textbf{1~290} \\ \bottomrule
  \end{tabular}
  \caption{Estimated workload to run the 2021--2022 edition of the introductory
    Python programming course for 442 students with 1 lecturer, 7 teaching
    assistants and 3 undergraduate students who serve as teaching assistants
    \citep{gordon_undergraduate_2013}.}\label{tab:workload}
\end{table}

\subsection{Learning analytics and educational data mining}

\begin{figure*}
  \centering
  \includegraphics[width=\linewidth]{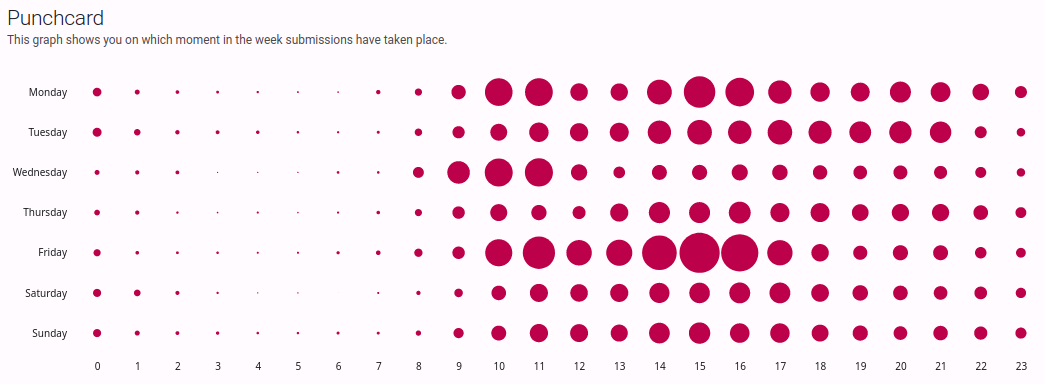}
  \caption{Punchcard from the \Dodona{} learning analytics page showing the
    distribution per weekday and per hour of all 331~734 solutions submitted
    during the 2021--2022 edition of the course (442
    students).}\label{fig:punchcard}
\end{figure*}

A longitudinal analysis of student submissions across the term shows that most
learning happens during the 13 weeks of educational activities and that students
don't have to catch up practicing their programming skills during the exam
period (Figure~\ref{fig:course-structure}). Active learning thus effectively
avoids procrastination. We observe that students submit solutions every day of
the week and show increased activity around hands-on sessions and in the run-up
to the weekly deadlines (Figure~\ref{fig:punchcard}). Weekends are also used to
work further on programming assignments, but students seem to be watching over a
good night's sleep.

\begin{figure*}
  \centering
  \includegraphics[width=0.7\linewidth]{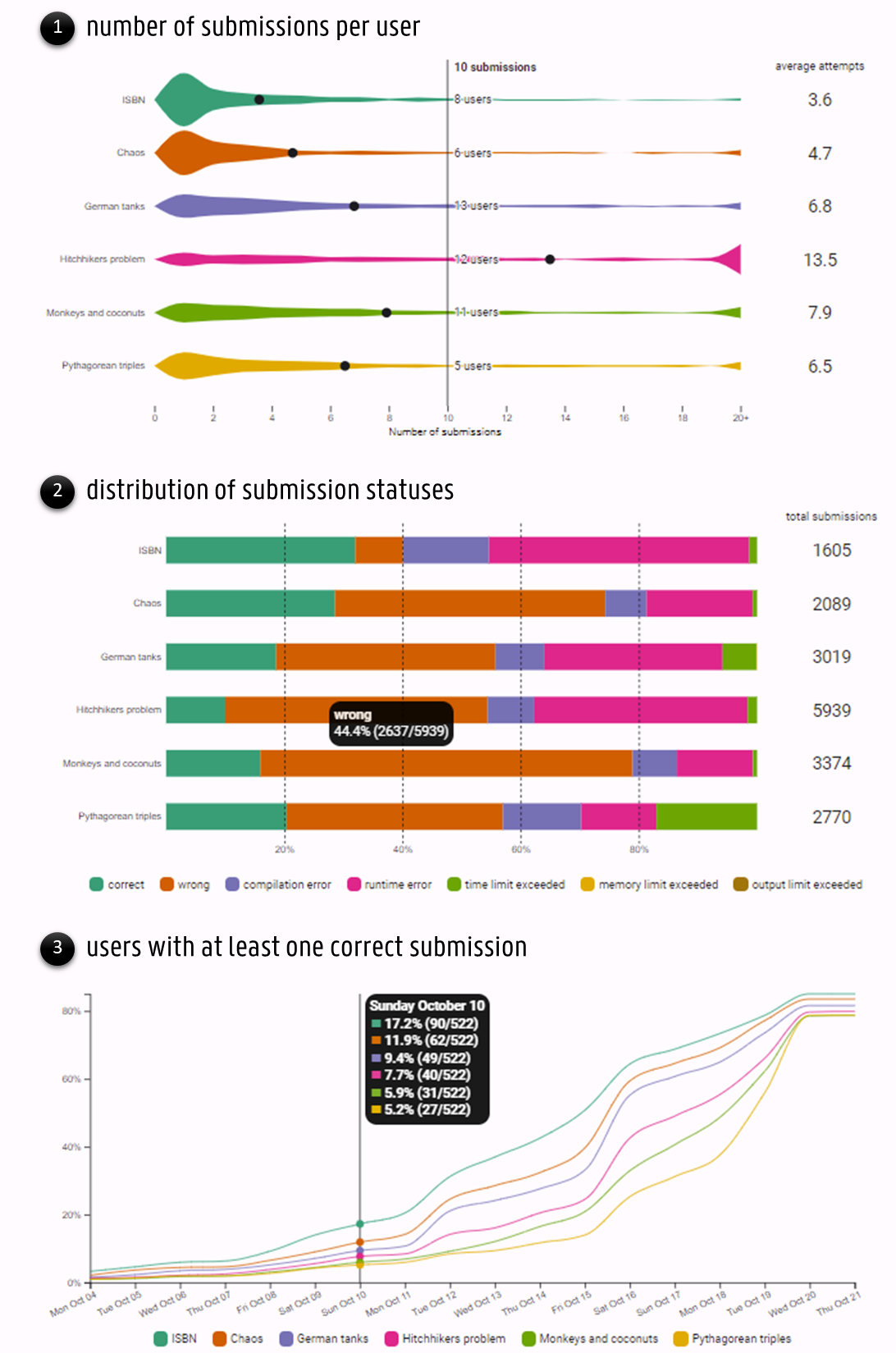}
  \caption{Interactive learning analytics on student submission behavior across
    programming assignments in the series where (unnested) loops are introduced
    in the course (2021--2022 edition). \textbf{Top}: Distribution of the number
    of student submissions per programming assignment. The larger the zone, the
    more students submitted a particular number of solutions. Black dot
    indicates the average number of submissions per student. \textbf{Middle}:
    Distribution of top-level submission statuses per programming
    assignment. \textbf{Bottom}: Progression over time of the percentage of
    students that correctly solved each assignment.}\label{fig:analytics}
\end{figure*}

Throughout a course edition, we use \Dodona{}'s series analytics to monitor how
students perform on our selection of programming assignments
(Figure~\ref{fig:analytics}). This allows us to make informed decisions and
appropriate interventions, for example when students experience issues with the
automated assessment configuration of a particular assignment or if the original
order of assignments in a series does not seem to align with our design goal to
present them in increasing order of difficulty. The first students that start
working on assignments usually are good performers. Seeing these early birds
having trouble with solving one of the assignments may give an early warning
that action is needed, as in improving the problem specification, adding extra
tips \& tricks, or better explaining certain programming concepts to all
students during lectures or hands-on sessions. Reversely, observing that many
students postpone working on their assignments until just before the deadline
might indicate that some assignments are simply too hard at this moment in time
through the learning pathway of the students or that completing the collection
of programming assignments interferes with the workload from other courses. Such
``deadline hugging'' patterns are also a good breeding ground for students to
resort on exchanging solutions among each other.

Using educational data mining techniques on historical data exported from
several editions of the course, we further investigated what aspects of
practicing programming skills promote or inhibit learning, or have no or minor
effect on the learning process \citep{van_petegem_passfail_2022}. It won't come
as a surprise that mid-term test scores are good predictors for a student's
final grade, because tests and exams are both summative assessments that are
organized and graded in the same way. However, we found that organizing a final
exam end-of-term is still a catalyst of learning, even for courses with a strong
focus of active learning during weeks of educational activities.

In evaluating if students gain deeper understanding when learning from their
mistakes while working progressively on their programming assignments, we found
the old adage that practice makes perfect to depend on what kind of mistakes
students make. Learning to code requires mastering two major competences:
\begin{enumerate*}[i)]
  \item getting familiar with the syntax and semantics of a programming language
    to express the steps for solving a problem in a formal way, so that the
    algorithm can be executed by a computer, and
  \item problem solving itself. It turns out that staying stuck longer on
    compilation errors (mistakes against the syntax of the programming language)
    inhibits learning, whereas taking progressively more time to get rid of
    logical errors (reflective of solving a problem with a wrong algorithm) as
    assignments get more complex actually promotes learning. After all, time
    spent in discovering solution strategies while thinking about logical errors
    can be reclaimed multifold when confronted with similar issues in later
    assignments \citep{glass_fewer_2022}.
\end{enumerate*}
These findings neatly align with the claim of \citet{edwards_separation_2018}
that problem solving is a higher-order learning task in Bloom's Taxonomy
(analysis and synthesis) than language syntax (knowledge, comprehension, and
application).

Using historical data from previous course editions, we can also make highly
accurate predictions about what students will pass or fail the current course
edition \citep{van_petegem_passfail_2022}. This can already be done after a few
weeks into the course, so remedial actions for at-risk students can be started
well in time. The approach is privacy-friendly as we only need to process
metadata on student submissions for programming assignments and results from
automated and manual assessment extracted from \Dodona{}. Given that cohort
sizes are large enough, historical data from a single course edition are already
enough to make accurate predictions.

\section{A broader perspective}\label{perspective}

\begin{figure*}
  \centering
  \includegraphics[width=\linewidth]{\Danon{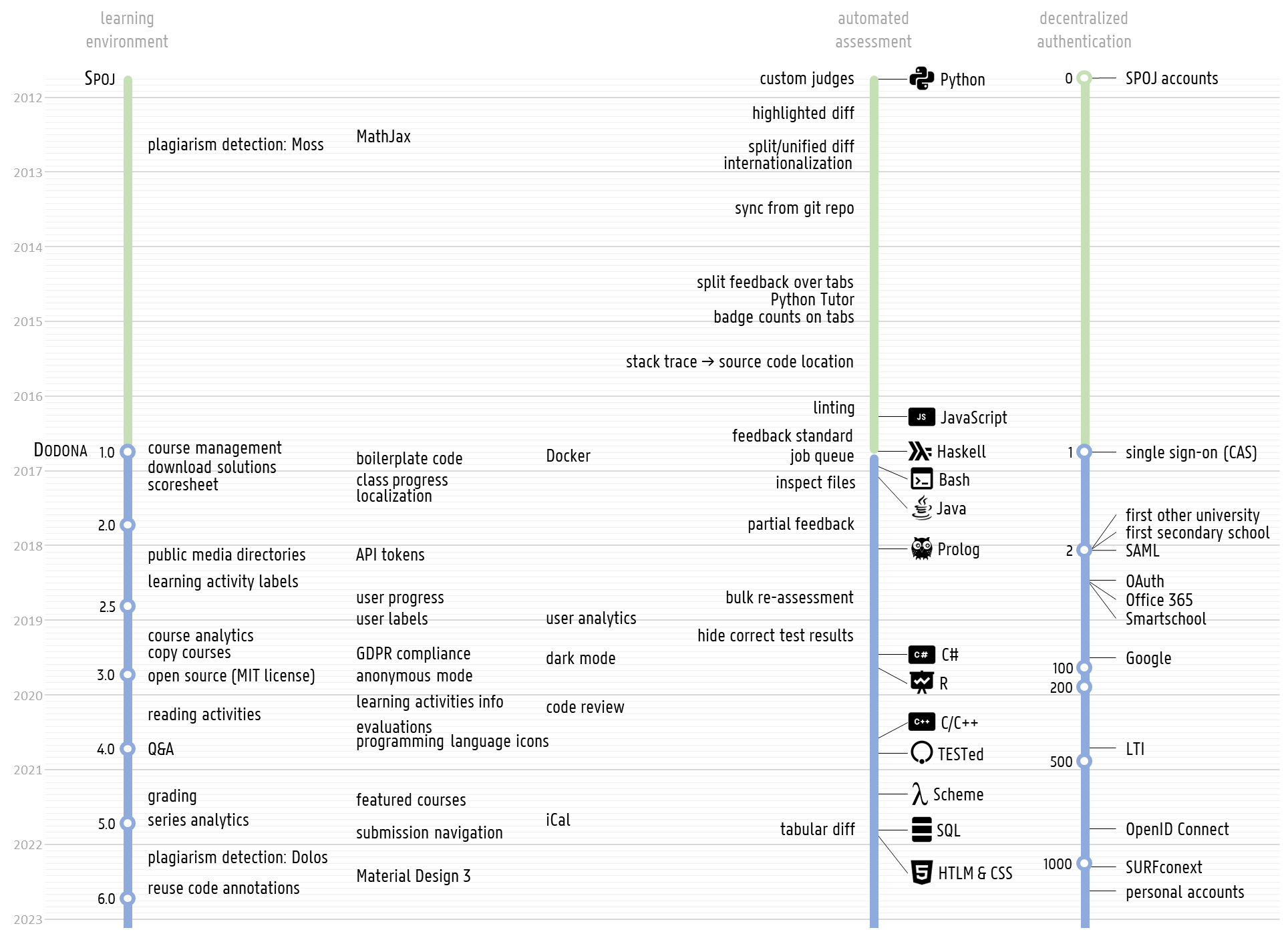}{history-anon.png}}
  \caption{Historical overview with milestones from running a Python course
    supported by automated assessment for eleven years and developing \Dodona{}
    for six years. Vertical positions of labels indicate when milestones were
    introduced. We used a Python judge in the Sphere Online Judge (SPOJ) to
    provide automated assessment during the first five years of the course,
    which was afterwards migrated to \Dodona{}. \textbf{From left to right}:
    features, judges and authentication-support introduced in
    SPOJ/\Dodona{}. \textbf{Numbers to the left of the tracks}: major \Dodona{}
    releases (learning environment), and number of institutions using \Dodona{}
    (decentralized authentication).}\label{fig:history}
\end{figure*}

\begin{figure*}
  \centering
  \includegraphics[width=0.45\linewidth]{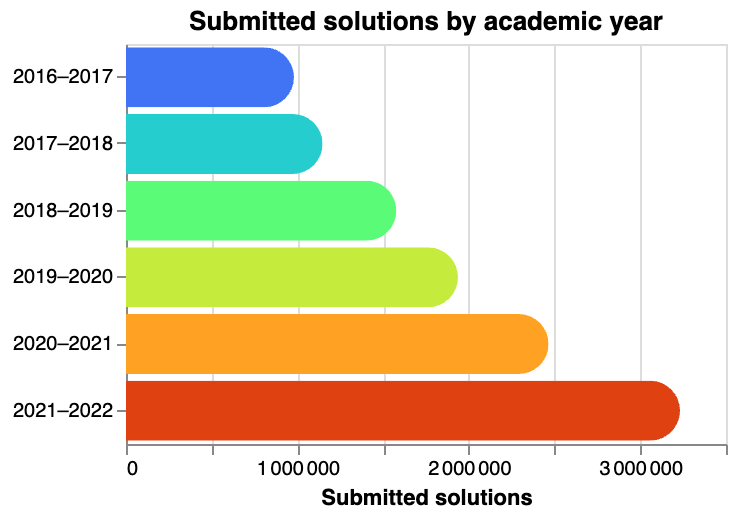}
  \includegraphics[width=0.45\linewidth]{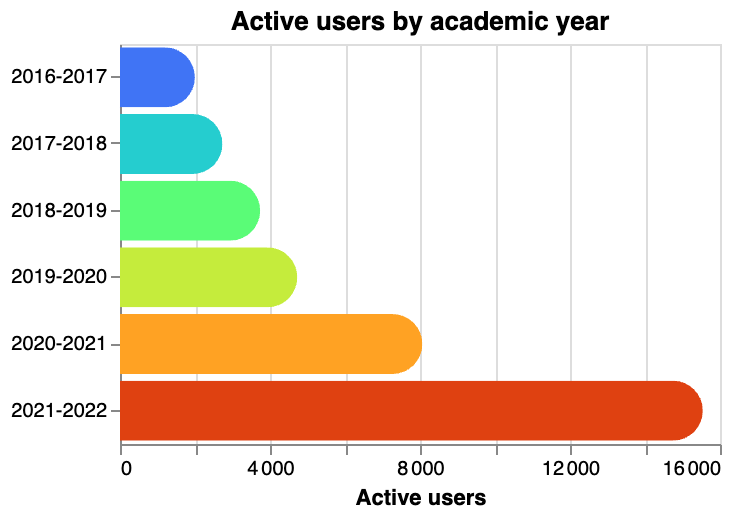}
  \caption{Overview of the number of submitted solutions and active users by
    academic year. Users were active when they submitted at least one solution
    for a programming assignment during the academic year.}\label{fig:adoption}
\end{figure*}

At the release of \Dodona{} version 6.0 (September 2022) --- eleven years after
we introduced automated assessment in our Python programming course and six
years after we started developing \Dodona{} --- the list of supported features
(Figure~\ref{fig:history}) and adoption beyond our own course
(Figure~\ref{fig:adoption}) have come a long way. The online learning platform
is now used in more than 1~000 schools, colleges and
universities\footnote{\Durl{https://dodona.ugent.be/en/support-us/}} mainly
across Flanders (Belgium) and the Netherlands, where 36 thousand students
altogether have submitted more than 11 million solutions for programming
assignments. Renewed interest in embedding computational thinking into formal
education has definitely been an important stimulator for such a broad adoption
\citep{wing_computational_2006}. The careful design choices and state-of-the art
software development that gave \Dodona{} the versatility, flexibility and
accessibility needed to lower the barrier for educational technology to find its
way beyond the context for which it is was initially created
\citep{rosling_enhancing_2008} also helped broaden adoption. This includes
single sign-on and trusted identities through decentralized authentication,
support for a wide variety of programming languages and assessment strategies
through a generic infrastructure for automated assessment, transparent
synchronization of content created in external git repositories,
computer-assisted code reviews and grading, plagiarism detection,
interoperability with learning management systems and IDEs. To stimulate
community building, we organize annual TeachMeets where teachers gather to share
ideas and good practices for teaching with \Dodona{}. These events spark new
ideas that we can integrate into \Dodona{} and that teachers can adopt in
designing their own courses.

\Dodona{} now houses a collection of 10 thousand reading activities and
programming assignments with support for automated assessment, which allows
teachers to compose learning paths that best fit their needs by mixing their own
learning activities with activities designed by others. Needless to say that
reuse of programming assignments is hampered by the fact that assignments target
different audiences, learning contexts and learning objectives. We however tried
to make content creation, publishing, and sharing as straightforward as
possible, and we constantly seek for new opportunities to make assignments more
FAIR\@: findable, accessible, interoperable, and reusable
\citep{wilkinson_fair_2016}. Effective information-seeking is currently
supported through faceted searching based on a flexible labeling scheme for
learning activities, featured courses, and rich metadata about learning
activities that teachers can consult on an information page containing author
attribution, general assessment configuration details, model solutions, usage
statistics and other background information provided by the authors.

The variety of programming languages used in education
\citep{crick_analysis_2017} is reflected by the many judges that have been
developed for \Dodona{}, ranging from general purpose programming and scripting
languages (Bash, C, C++, C\#, Haskell, Java, JavaScript, Prolog, Python, Scheme)
to languages dedicated to data science and statistics (R)
\citep{nust_rockerverse_2020}, databases (SQL) and web design (HTML \& CSS). We
can make three observations from educational software testing to provide
automated feedback on solutions that students submit for programming
assignments. First, implementing a judge takes time and careful design. Setting
up an initial prototype is easily done, which explains why there are so many
(industrial) software testing frameworks around. But then it becomes more
challenging to make the judge generate rich feedback that supports student
learning in the best possible way, while keeping assignment-specific assessment
configurations as lightweight as possible by supporting many out-of-the-box test
strategies (``fat judge / slim test suite''). To avoid frustrated students and
teachers, the performance of judges in educational practice must also bring
top-quality in terms of speed, robustness and security
\citep{peveler_comparing_2019}. Second, different educational judges share a lot
of supported software testing strategies that are merely re-implemented to
target different programming languages
\citep{ala-mutka_survey_2005,caiza_programming_2013,douce_automatic_2005,ihantola_review_2010,paiva_automated_2022,wasik_survey_2018,wilcox_testing_2016}. Third,
teachers often reuse programming assignments across different programming
languages. This may require modifying the problem description and/or creating a
separate assessment configuration for each target language.

To make the design of assignments with support for automated assessment less
dependent on a specific programming language, \Dodona{} also provides a judge
based on TESTed: an educational test framework that unifies the specification of
software tests across programming languages \citep{strijbol_tested_nodate}. This
directly benefits educators: assignments can be used across programming
languages, designing assignments for different programming languages can be done
using the same test framework, and the cost of providing automated assessment
for new programming languages is dramatically reduced as TESTed implements the
core components of automated assessment in a generic language-agnostic way. It
only needs a thin layer of language specific configurations for each individual
programming language and currently supports C, Haskell, Java, JavaScript, Kotlin
and Python. \Dodona{} itself also takes up some responsibilities for automated
assessment through its generic infrastructure: provisioning a secure runtime
environment and presenting feedback that is expressed in a standard way. At the
same time, this infrastructure is not very restrictive as to how automated
assessment can be performed.

\section{Conclusions}\label{conclusions}

The formative assessment model of Nicol and Macfarlane-Dick and their seven
principles of good practice feedback \citep{nicol_formative_2006} align very
well with how \Dodona{} facilitates students in self-regulated learning while
practicing their programming skills and how it provides information to teachers
that helps to shape their teaching. For students, active learning promises to
reinforce learning by creating shorter feedback loops that help them make
adjustments early on in the learning process. But scaling up feedback
provisioning throughout the entire learning process might become a real
bottleneck for teachers and instructors. \Dodona{} therefore aims at saving
valuable teacher time for maintaining a collaborative and responsive dialogue
with students based on high-quality and timely feedback. However, while it may
be the ultimate ideal for some, current educational technology does not yet
allow to completely automate the entire feedback loop. Supporting the human
aspect of learning and teaching is therefore an important focus in designing the
\Dodona{} user experience. Students can track and remedy potential errors in
their code with a built-in graphical debugger, ask online questions directly on
their submitted solutions with the integrated Q\&A module, and monitor their own
progress from learning analytics dashboards. Teachers can customize learning
paths with their own learning materials and interactive assignments, share
materials with their colleagues, monitor student progress (individually or in
group) using learning analytics dashboards, organize high-stakes tests and exams
with automated feedback, assess students with rich feedback using a grading
module with support for code reviews, and detect and prevent plagiarism with
dedicated and interactive tools.

Pushing the boundaries of \Dodona{} as a virtual co-teacher that becomes
gradually smarter at supporting or automating pedagogical tasks is an active
area of our research. Observing how \Dodona{} inspired so many colleagues to
increasingly bring active and blended learning into their educational practice
keeps broadening our vision and is in itself an inspiration for novel
features. If reading this paper also triggered your curiosity to start exploring
the platform, feel free to create an account and request teacher
rights\footnote{\Durl{https://dodona.ugent.be/en/rights\_requests/new/}} that
allow you to set up learning paths for your own courses and create your own
programming assignments with support for automated assessment. If you are more
tech savvy, you can also develop your own judges to support automated assessment
according to your own pedagogical vision. Sharing these learning materials with
colleagues is one way to contribute to the platform. In addition, the source
code of \Dodona{} is made publicly available on
GitHub\footnote{\Durl{https://github.com/dodona-edu}}, where we welcome bug
reports and feature requests documented as issues, user experiences shared as
discussions, and contributions submitted as pull requests. Over the last six
years \Dodona{} has traveled wherever its code has taken us, so wherever you go,
may the source be with you.

\section*{Acknowledgments}

\Danon{
  We are grateful for the financial support of Ghent University (Belgium) and
  the Flemish Government (Belgium, Voorsprongfonds) through numerous innovation
  in education grants. Part of this work was also supported by the Research
  Foundation --- Flanders (FWO) for ELIXIR Belgium (I002819N). Thanks to Ghent
  University (Belgium) for granting us the 2018 Minerva Award for our
  contributions to active learning and innovation in education through the
  development of Dodona. Thanks to the Flemish Government (Belgium) for granting
  us with the 2022 Flanders Digital Award for providing each student high
  quality education through Dodona. Thanks to Johan Van Camp and his team at the
  Ghent University Data Center for hosting Dodona software services. Thanks to
  Hanne Elsen (UGent Data Protection Office) for assisting us with GDPR and
  privacy-related issues. Thanks to the computer science students and
  instructors who helped in developing the Dodona platform as interns, while
  working on their master's thesis or while running courses: Winnie De Ridder,
  Tibo D'hondt, Lucianos Lionakis, Felix Van der Jeugt, Mathieu Coussens, Pieter
  De Clercq, Timon De Backer, Ilion Beyst, Dieter Mourisse, Brecht Willems,
  Robbert Gurdeep Singh, Louise Deconick, Tim Ramlot, Bram Devlaminck, Freya Van
  Speybroeck, Toon Baeyens, Jeroen Tiebout and Anton Kindt. Thanks to all judge
  developers: Peter Dawyndt, Dieter Mourisse, Niels Neirynck, Felix Van der
  Jeugt, Charlotte Van Petegem, Bart Mesuere (Python); Peter Dawyndt, Dieter
  Mourisse, Bart Mesuere, Rien Maertens, Charlotte Van Petegem (JavaScript);
  Felix Van der Jeugt, Christophe Scholliers, Charlotte Van Petegem, Rien
  Maertens (Haskell); Niels Neirinck, Pieter Verschaffelt, Charlotte Van Petegem
  (Bash); Felix Van der Jeugt, Pieter De Clercq, Bart Mesuere, J. Steegmans
  (Java); Robbert Gurdeep Singh, Charlotte Van Petegem, Rien Maertens (Prolog);
  Dieter Mourisse (C\#); Charlotte Van Petegem, Viktor Verstraelen, Gust
  Bogaert, Koen Plevoets, Bart Mesuere (R); Maarten Vandercammen, Elisa Gonzalez
  Boix (C/C++); Boris Sels, Niko Strijbol, Charlotte Van Petegem, Peter Dawyndt,
  Bart Mesuere (TESTed); Mathijs Saey (Scheme); Stijn De Clercq, Quinten
  Vervynck, Brecht Willems (HTML \& CSS); Brecht Willems, Tim Ramlot, Pieter De
  Clercq (SQL); Thanks to Pieter De Clercq and Tobiah Lissens for developing the
  JetBrains IDE plugin, to Stijn De Clercq and Pieter De Clercq for developing
  the Visual Studio Code plugin and to Mathijs Saey at the Vrije Universiteit
  Brussel (Brussels, Belgium) for developing the DrRacket plugin. Thanks to all
  teachers and instructors who developed learning activities for Dodona and
  shared them on the platform as open educational resources. Thanks to all users
  who reported issues and provided feedback.
}{Acknowledgments are left out to make sure the paper is anonymous.}

\bibliography{bibliography.bib}

\begin{thebibliography}{}

\bibitem[Ak{\c c}ay{\i}r and Ak{\c c}ay{\i}r, 2018]{akcayir_flipped_2018}
Ak{\c c}ay{\i}r, G. and Ak{\c c}ay{\i}r, M. (2018).
\newblock The flipped classroom: {A} review of its advantages and challenges.
\newblock {\em Computers \& Education}, 126:334--345.

\bibitem[Ala-Mutka, 2005]{ala-mutka_survey_2005}
Ala-Mutka, K.~M. (2005).
\newblock A {Survey} of {Automated} {Assessment} {Approaches} for {Programming}
  {Assignments}.
\newblock {\em Computer Science Education}, 15(2):83--102.
\newblock Publisher: Routledge \_eprint:
  https://doi.org/10.1080/08993400500150747.

\bibitem[Alfadel et~al., 2021]{alfadel_use_2021}
Alfadel, M., Costa, D.~E., Shihab, E., and Mkhallalati, M. (2021).
\newblock On the {Use} of {Dependabot} {Security} {Pull} {Requests}.
\newblock In {\em 2021 {IEEE}/{ACM} 18th {International} {Conference} on
  {Mining} {Software} {Repositories} ({MSR})}, pages 254--265.
\newblock ISSN: 2574-3864.

\bibitem[Baker and Yacef, 2009]{baker_state_2009}
Baker, R. S. J.~d. and Yacef, K. (2009).
\newblock The {State} of {Educational} {Data} {Mining} in 2009: {A} {Review}
  and {Future} {Visions}.
\newblock {\em Journal of Educational Data Mining}, 1(1):3--17.
\newblock Number: 1.

\bibitem[Becker et~al., 2019]{becker_compiler_2019}
Becker, B.~A., Denny, P., Pettit, R., Bouchard, D., Bouvier, D.~J., Harrington,
  B., Kamil, A., Karkare, A., McDonald, C., Osera, P.-M., Pearce, J.~L., and
  Prather, J. (2019).
\newblock Compiler {Error} {Messages} {Considered} {Unhelpful}: {The}
  {Landscape} of {Text}-{Based} {Programming} {Error} {Message} {Research}.
\newblock In {\em Proceedings of the {Working} {Group} {Reports} on
  {Innovation} and {Technology} in {Computer} {Science} {Education}},
  {ITiCSE}-{WGR} '19, pages 177--210, New York, NY, USA. Association for
  Computing Machinery.

\bibitem[Bell, 2011]{bell_connectivism_2011}
Bell, F. (2011).
\newblock Connectivism: {Its} {Place} in {Theory}-{Informed} {Research} and
  {Innovation} in {Technology}-{Enabled} {Learning}.
\newblock {\em International Review of Research in Open and Distributed
  Learning}, 12(3):98--118.
\newblock Publisher: Athabasca University Press (AU Press).

\bibitem[Bishop and Verleger, 2013]{bishop_flipped_2013}
Bishop, J. and Verleger, M.~A. (2013).
\newblock The {Flipped} {Classroom}: {A} {Survey} of the {Research}.
\newblock pages 23.1200.1--23.1200.18.
\newblock ISSN: 2153-5965.

\bibitem[Bonar and Cunningham, 1988]{bonar_bridge_1988}
Bonar, J.~G. and Cunningham, R. (1988).
\newblock Bridge: {Intelligent} tutoring with intermediate representations.
\newblock Technical report, CARNEGIE-MELLON UNIV PITTSBURGH PA ARTIFICIAL
  INTELLIGENCE AND PSYCHOLOGY.

\bibitem[Brooks and Kugler, 1987]{brooks_no_1987}
Brooks, F. and Kugler, H. (1987).
\newblock {\em No silver bullet}.
\newblock April.

\bibitem[Caiza and {\'A}lamo~Ramiro, 2013]{caiza_programming_2013}
Caiza, J.~C. and {\'A}lamo~Ramiro, J. M.~d. (2013).
\newblock Programming assignments automatic grading: review of tools and
  implementations.
\newblock In {\em 7th {International} {Technology}, {Education} and
  {Development} {Conference} ({INTED2013}) {\textbar} 7th {International}
  {Technology}, {Education} and {Development} {Conference} ({INTED2013})
  {\textbar} 04/03/2013 - 06/03/2013 {\textbar} {Valencia}, {Spain}}, pages
  5691--5700, Valencia, Spain. E.T.S.I. Telecomunicaci{\'o}n (UPM).
\newblock Num Pages: 10.

\bibitem[Caswell et~al., 2008]{caswell_open_2008}
Caswell, T., Henson, S., Jensen, M., and Wiley, D. (2008).
\newblock Open {Educational} {Resources}: {Enabling} universal education.
\newblock {\em International Review of Research in Open and Distributed
  Learning}, 9(1):1--11.
\newblock Publisher: Athabasca University Press (AU Press).

\bibitem[Cervone, 2012]{cervone_mathjax_2012}
Cervone, D. (2012).
\newblock {MathJax}: a platform for mathematics on the {Web}.
\newblock {\em Notices of the AMS}, 59(2):312--316.

\bibitem[Cheang et~al., 2003]{cheang_automated_2003}
Cheang, B., Kurnia, A., Lim, A., and Oon, W.-C. (2003).
\newblock On automated grading of programming assignments in an academic
  institution.
\newblock {\em Computers \& Education}, 41(2):121--131.

\bibitem[Chickering and Gamson, 1987]{chickering_seven_1987}
Chickering, A.~W. and Gamson, Z.~F. (1987).
\newblock Seven {Principles} for {Good} {Practice} in {Undergraduate}
  {Education}.
\newblock {\em AAHE Bulletin}.

\bibitem[Cooper, 2000]{cooper_facilitating_2000}
Cooper, N.~J. (2000).
\newblock Facilitating {Learning} from {Formative} {Feedback} in {Level} 3
  {Assessment}.
\newblock {\em Assessment \& Evaluation in Higher Education}, 25(3):279--291.
\newblock Publisher: Routledge \_eprint: https://doi.org/10.1080/713611435.

\bibitem[Crick, 2017]{crick_analysis_2017}
Crick, T. (2017).
\newblock An {Analysis} of {Introductory} {Programming} {Courses} at {UK}
  {Universities}.
\newblock {\em The Art, Science, and Engineering of Programming}, 1(2).

\bibitem[Dawson, 2017]{dawson_assessment_2017}
Dawson, P. (2017).
\newblock Assessment rubrics: towards clearer and more replicable design,
  research and practice.
\newblock {\em Assessment \& Evaluation in Higher Education}, 42(3):347--360.
\newblock Publisher: Routledge \_eprint:
  https://doi.org/10.1080/02602938.2015.1111294.

\bibitem[Dooley, 2011]{dooley_software_2011}
Dooley, J. (2011).
\newblock {\em Software {Development} and {Professional} {Practice}}.
\newblock Apress, Berkeley, CA.

\bibitem[Douce et~al., 2005]{douce_automatic_2005}
Douce, C., Livingstone, D., and Orwell, J. (2005).
\newblock Automatic test-based assessment of programming: {A} review.
\newblock {\em Journal on Educational Resources in Computing}, 5(3):4--es.

\bibitem[Downes, 2007]{downes_models_2007}
Downes, S. (2007).
\newblock Models for {Sustainable} {Open} {Educational} {Resources}.
\newblock {\em Interdisciplinary Journal of E-Learning and Learning Objects},
  3(1):29--44.
\newblock Publisher: Informing Science Institute.

\bibitem[Edwards et~al., 2018]{edwards_separation_2018}
Edwards, J.~M., Fulton, E.~K., Holmes, J.~D., Valentin, J.~L., Beard, D.~V.,
  and Parker, K.~R. (2018).
\newblock Separation of syntax and problem solving in {Introductory} {Computer}
  {Programming}.
\newblock In {\em 2018 {IEEE} {Frontiers} in {Education} {Conference} ({FIE})},
  pages 1--5.
\newblock ISSN: 2377-634X.

\bibitem[Edwards, 2004]{edwards_using_2004}
Edwards, S.~H. (2004).
\newblock Using software testing to move students from trial-and-error to
  reflection-in-action.
\newblock In {\em Proceedings of the 35th {SIGCSE} technical symposium on
  {Computer} science education}, {SIGCSE} '04, pages 26--30, New York, NY, USA.
  Association for Computing Machinery.

\bibitem[Farrell et~al., 2002]{farrell_assertions_2002}
Farrell, S., Reid, I., Orchard, D., Sankar, K., Moses, T., Edwards, E.~N.,
  Pato, J., Knouse, C., Cantor, O.~S., and Platt, D. (2002).
\newblock Assertions and {Protocol} for the {OASIS} {Security} {Assertion}
  {Markup} {Language} ({SAML}).
\newblock {\em Organization for the Advancement of Structured Information
  Standards (OASIS) Standard (November 2002), http://www. oasis-open.
  org/committees/download. php/1371/oasis-sstc-saml-core-1.0. pdf}.

\bibitem[Ferguson, 2012]{ferguson_learning_2012}
Ferguson, R. (2012).
\newblock Learning analytics: drivers, developments and challenges.
\newblock {\em International Journal of Technology Enhanced Learning},
  4(5/6):304--317.
\newblock Number: 5/6.

\bibitem[Fincher, 1999]{fincher_what_1999}
Fincher, S. (1999).
\newblock What are we doing when we teach programming?
\newblock In {\em {FIE}'99 {Frontiers} in {Education}. 29th {Annual}
  {Frontiers} in {Education} {Conference}. {Designing} the {Future} of
  {Science} and {Engineering} {Education}. {Conference} {Proceedings} ({IEEE}
  {Cat}. {No}.{99CH37011}}, volume~1, pages 12A4/1--12A4/5 vol.1.
\newblock ISSN: 0190-5848.

\bibitem[Forisek, 2006]{forisek_suitability_2006}
Forisek, M. (2006).
\newblock On the {Suitability} of {Programming} {Tasks} for {Automated}
  {Evaluation}.
\newblock {\em Informatics in Education}, 5(1):63--76.
\newblock Publisher: Vilnius University Institute of Data Science and Digital
  Technologies.

\bibitem[Gibbs and Simpson, 2005]{gibbs_conditions_2005}
Gibbs, G. and Simpson, C. (2005).
\newblock Conditions {Under} {Which} {Assessment} {Supports}
  {Students}{\textquoteright} {Learning}.
\newblock {\em Learning and Teaching in Higher Education}, (1):3--31.
\newblock Number: 1 Publisher: University of Gloucestershire.

\bibitem[Glass and Kang, 2022]{glass_fewer_2022}
Glass, A.~L. and Kang, M. (2022).
\newblock Fewer students are benefiting from doing their homework: an
  eleven-year study.
\newblock {\em Educational Psychology}, 42(2):185--199.
\newblock Publisher: Routledge \_eprint:
  https://doi.org/10.1080/01443410.2020.1802645.

\bibitem[Gordon et~al., 2013]{gordon_undergraduate_2013}
Gordon, J., Henry, P., and Dempster, M. (2013).
\newblock Undergraduate {Teaching} {Assistants}: {A} {Learner}-{Centered}
  {Model} for {Enhancing} {Student} {Engagement} in the {First}-{Year}
  {Experience}.
\newblock {\em International Journal of Teaching and Learning in Higher
  Education}, 25(1):103--109.
\newblock Publisher: International Society for Exploring Teaching and Learning.

\bibitem[Graham et~al., 2021]{graham_foundations_2021}
Graham, D., Black, R., and Veenendaal, E.~v. (2021).
\newblock {\em Foundations of {Software} {Testing} {ISTQB} {Certification}, 4th
  edition}.
\newblock Cengage Learning.
\newblock Google-Books-ID: mOwxEAAAQBAJ.

\bibitem[Guo, 2013]{guo_online_2013}
Guo, P.~J. (2013).
\newblock Online python tutor: embeddable web-based program visualization for
  cs education.
\newblock In {\em Proceeding of the 44th {ACM} technical symposium on
  {Computer} science education}, {SIGCSE} '13, pages 579--584, New York, NY,
  USA. Association for Computing Machinery.

\bibitem[Hanks et~al., 2011]{hanks_pair_2011}
Hanks, B., Fitzgerald, S., McCauley, R., Murphy, L., and Zander, C. (2011).
\newblock Pair programming in education: a literature review.
\newblock {\em Computer Science Education}, 21(2):135--173.
\newblock Publisher: Routledge \_eprint:
  https://doi.org/10.1080/08993408.2011.579808.

\bibitem[Hardt, 2012]{hardt_oauth_2012}
Hardt, D. (2012).
\newblock The {OAuth} 2.0 {Authorization} {Framework}.
\newblock Request for {Comments} RFC 6749, Internet Engineering Task Force.
\newblock Num Pages: 76.

\bibitem[Hattie and Timperley, 2007]{hattie_power_2007}
Hattie, J. and Timperley, H. (2007).
\newblock The {Power} of {Feedback}.
\newblock {\em Review of Educational Research}, 77(1):81--112.
\newblock Publisher: American Educational Research Association.

\bibitem[Hollingsworth, 1960]{hollingsworth_automatic_1960}
Hollingsworth, J. (1960).
\newblock Automatic graders for programming classes.
\newblock {\em Communications of the ACM}, 3(10):528--529.

\bibitem[Hunt, 1999]{hunt_pragmatic_1999}
Hunt, A. (1999).
\newblock {\em The pragmatic programmer}.
\newblock Pearson Education India.

\bibitem[Hyl{\'e}n, 2021]{hylen_open_2021}
Hyl{\'e}n, J. (2021).
\newblock Open educational resources: {Opportunities} and challenges.
\newblock Publisher: OECD.

\bibitem[Ihantola et~al., 2010]{ihantola_review_2010}
Ihantola, P., Ahoniemi, T., Karavirta, V., and Sepp{\"a}l{\"a}, O. (2010).
\newblock Review of recent systems for automatic assessment of programming
  assignments.
\newblock In {\em Proceedings of the 10th {Koli} {Calling} {International}
  {Conference} on {Computing} {Education} {Research}}, Koli {Calling} '10,
  pages 86--93, New York, NY, USA. Association for Computing Machinery.

\bibitem[Jackson, 2000]{jackson_semi-automated_2000}
Jackson, D. (2000).
\newblock A semi-automated approach to online assessment.
\newblock {\em ACM SIGCSE Bulletin}, 32(3):164--167.

\bibitem[Jackson and Usher, 1997]{jackson_grading_1997}
Jackson, D. and Usher, M. (1997).
\newblock Grading student programs using {ASSYST}.
\newblock In {\em Proceedings of the twenty-eighth {SIGCSE} technical symposium
  on {Computer} science education}, {SIGCSE} '97, pages 335--339, New York, NY,
  USA. Association for Computing Machinery.

\bibitem[Karabenick and Knapp, 1991]{karabenick_relationship_1991}
Karabenick, S.~A. and Knapp, J.~R. (1991).
\newblock Relationship of academic help seeking to the use of learning
  strategies and other instrumental achievement behavior in college students.
\newblock {\em Journal of Educational Psychology}, 83(2):221--230.
\newblock Place: US Publisher: American Psychological Association.

\bibitem[Kelleher et~al., 2002]{kelleher_alice2_2002}
Kelleher, C., Cosgrove, D., Culyba, D., Forlines, C., Pratt, J., and Pausch, R.
  (2002).
\newblock Alice2: programming without syntax errors.
\newblock In {\em User {Interface} {Software} and {Technology}}, volume~2,
  pages 35--36. Citeseer.
\newblock Issue: 2.

\bibitem[Keuning et~al., 2018]{keuning_systematic_2018}
Keuning, H., Jeuring, J., and Heeren, B. (2018).
\newblock A {Systematic} {Literature} {Review} of {Automated} {Feedback}
  {Generation} for {Programming} {Exercises}.
\newblock {\em ACM Transactions on Computing Education}, 19(1):3:1--3:43.

\bibitem[Kosowski et~al., 2008]{kosowski_application_2008}
Kosowski, A., Ma{\l }afiejski, M., and Noi{\'n}ski, T. (2008).
\newblock Application of an {Online} {Judge} \& {Contester} {System} in
  {Academic} {Tuition}.
\newblock In Leung, H., Li, F., Lau, R., and Li, Q., editors, {\em Advances in
  {Web} {Based} {Learning} {\textendash} {ICWL} 2007}, Lecture {Notes} in
  {Computer} {Science}, pages 343--354, Berlin, Heidelberg. Springer.

\bibitem[Laplante, 2007]{laplante_what_2007}
Laplante, P.~A. (2007).
\newblock {\em What every engineer should know about software engineering}.
\newblock CRC Press.

\bibitem[Lebuda and Karwowski, 2013]{lebuda_tell_2013}
Lebuda, I. and Karwowski, M. (2013).
\newblock Tell {Me} {Your} {Name} and {I}'ll {Tell} {You} {How} {Creative}
  {Your} {Work} {Is}: {Author}'s {Name} and {Gender} as {Factors} {Influencing}
  {Assessment} of {Products}' {Creativity} in {Four} {Different} {Domains}.
\newblock {\em Creativity Research Journal}, 25(1):137--142.
\newblock Publisher: Routledge \_eprint:
  https://doi.org/10.1080/10400419.2013.752297.

\bibitem[Leiba, 2012]{leiba_oauth_2012}
Leiba, B. (2012).
\newblock {OAuth} {Web} {Authorization} {Protocol}.
\newblock {\em IEEE Internet Computing}, 16(1):74--77.
\newblock Conference Name: IEEE Internet Computing.

\bibitem[Maertens et~al., 2022]{maertens_dolos_2022}
Maertens, R., Van~Petegem, C., Strijbol, N., Baeyens, T., Jacobs, A.~C.,
  Dawyndt, P., and Mesuere, B. (2022).
\newblock Dolos: {Language}-agnostic plagiarism detection in source code.
\newblock {\em Journal of Computer Assisted Learning}, n/a(n/a).
\newblock \_eprint: https://onlinelibrary.wiley.com/doi/pdf/10.1111/jcal.12662.

\bibitem[Malouff et~al., 2013]{malouff_risk_2013}
Malouff, J.~M., Emmerton, A.~J., and Schutte, N.~S. (2013).
\newblock The {Risk} of a {Halo} {Bias} as a {Reason} to {Keep} {Students}
  {Anonymous} {During} {Grading}.
\newblock {\em Teaching of Psychology}, 40(3):233--237.
\newblock Publisher: SAGE Publications Inc.

\bibitem[Malouff and Thorsteinsson, 2016]{malouff_bias_2016}
Malouff, J.~M. and Thorsteinsson, E.~B. (2016).
\newblock Bias in grading: {A} meta-analysis of experimental research findings.
\newblock {\em Australian Journal of Education}, 60(3):245--256.
\newblock Publisher: SAGE Publications Ltd.

\bibitem[Mani et~al., 2014]{mani_better_2014}
Mani, A., Venkataramani, D., Petit~Silvestre, J., and Roura~Ferret, S. (2014).
\newblock Better feedback for educational online judges.
\newblock pages 176--183. SciTePress.
\newblock Accepted: 2015-06-04T08:47:03Z.

\bibitem[Marrero and Settle, 2005]{marrero_testing_2005}
Marrero, W. and Settle, A. (2005).
\newblock Testing first: emphasizing testing in early programming courses.
\newblock In {\em Proceedings of the 10th annual {SIGCSE} conference on
  {Innovation} and technology in computer science education}, {ITiCSE} '05,
  pages 4--8, New York, NY, USA. Association for Computing Machinery.

\bibitem[McCabe, 1976]{mccabe_complexity_1976}
McCabe, T. (1976).
\newblock A {Complexity} {Measure}.
\newblock {\em IEEE Transactions on Software Engineering}, SE-2(4):308--320.
\newblock Conference Name: IEEE Transactions on Software Engineering.

\bibitem[Miller and Maloney, 1963]{miller_systematic_1963}
Miller, J.~C. and Maloney, C.~J. (1963).
\newblock Systematic mistake analysis of digital computer programs.
\newblock {\em Communications of the ACM}, 6(2):58--63.
\newblock Publisher: ACM New York, NY, USA.

\bibitem[Myers, 1986]{myers_anond_1986}
Myers, E.~W. (1986).
\newblock {AnO}({ND}) difference algorithm and its variations.
\newblock {\em Algorithmica}, 1(1):251--266.

\bibitem[Nandi et~al., 2012]{nandi_evaluating_2012}
Nandi, D., Hamilton, M., and Harland, J. (2012).
\newblock Evaluating the quality of interaction in asynchronous discussion
  forums in fully online courses.
\newblock {\em Distance Education}, 33(1):5--30.
\newblock Publisher: Routledge \_eprint:
  https://doi.org/10.1080/01587919.2012.667957.

\bibitem[Newman and Schwager, 1993]{newman_students_1993}
Newman, R.~S. and Schwager, M.~T. (1993).
\newblock Students' {Perceptions} of the {Teacher} and {Classmates} in
  {Relation} to {Reported} {Help} {Seeking} in {Math} {Class}.
\newblock {\em The Elementary School Journal}, 94(1):3--17.
\newblock Publisher: The University of Chicago Press.

\bibitem[Nicol and Macfarlane-Dick, 2006]{nicol_formative_2006}
Nicol, D.~J. and Macfarlane-Dick, D. (2006).
\newblock Formative assessment and self-regulated learning: a model and seven
  principles of good feedback practice.
\newblock {\em Studies in Higher Education}, 31(2):199--218.
\newblock Publisher: Routledge \_eprint:
  https://doi.org/10.1080/03075070600572090.

\bibitem[Nidhra and Dondeti, 2012]{nidhra_black_2012}
Nidhra, S. and Dondeti, J. (2012).
\newblock Black box and white box testing techniques-a literature review.
\newblock {\em International Journal of Embedded Systems and Applications
  (IJESA)}, 2(2):29--50.

\bibitem[N{\"u}st et~al., 2020]{nust_rockerverse_2020}
N{\"u}st, D., Eddelbuettel, D., Bennett, D., Cannoodt, R., Clark, D.,
  Dar{\'o}czi, G., Edmondson, M., Fay, C., Hughes, E., Kjeldgaard, L., Lopp,
  S., Marwick, B., Nolis, H., Nolis, J., Ooi, H., Ram, K., Ross, N., Shepherd,
  L., S{\'o}lymos, P., Swetnam, T.~L., Turaga, N., Van~Petegem, C., Williams,
  J., Willis, C., and Xiao, N. (2020).
\newblock The {Rockerverse}: {Packages} and {Applications} for
  {Containerisation} with {R}.
\newblock {\em The R Journal}, 12(1):437--461.

\bibitem[Oberkampf and Roy, 2010]{oberkampf_verification_2010}
Oberkampf, W.~L. and Roy, C.~J. (2010).
\newblock {\em Verification and {Validation} in {Scientific} {Computing}}.
\newblock Cambridge University Press.
\newblock Google-Books-ID: 7d26zLEJ1FUC.

\bibitem[Paiva et~al., 2022]{paiva_automated_2022}
Paiva, J.~C., Leal, J.~P., and Figueira, {\'A}. (2022).
\newblock Automated {Assessment} in {Computer} {Science} {Education}: {A}
  {State}-of-the-{Art} {Review}.
\newblock {\em ACM Transactions on Computing Education}, 22(3):34:1--34:40.

\bibitem[{Peter Norvig}, 2001]{peter_norvig_teach_2001}
{Peter Norvig} (2001).
\newblock Teach {Yourself} {Programming} in {Ten} {Years}.

\bibitem[Peveler et~al., 2019]{peveler_comparing_2019}
Peveler, M., Maicus, E., and Cutler, B. (2019).
\newblock Comparing {Jailed} {Sandboxes} vs {Containers} {Within} an
  {Autograding} {System}.
\newblock In {\em Proceedings of the 50th {ACM} {Technical} {Symposium} on
  {Computer} {Science} {Education}}, {SIGCSE} '19, pages 139--145, New York,
  NY, USA. Association for Computing Machinery.

\bibitem[Pintrich, 1995]{pintrich_understanding_1995}
Pintrich, P.~R. (1995).
\newblock Understanding self-regulated learning.
\newblock {\em New Directions for Teaching and Learning}, 1995(63):3--12.
\newblock \_eprint:
  https://onlinelibrary.wiley.com/doi/pdf/10.1002/tl.37219956304.

\bibitem[Popham, 1997]{popham_whats_1997}
Popham, W.~J. (1997).
\newblock What's {Wrong}--and {What}'s {Right}--with {Rubrics}.
\newblock {\em Educational Leadership}, 55(2):72--75.

\bibitem[Prince, 2004]{prince_does_2004}
Prince, M. (2004).
\newblock Does {Active} {Learning} {Work}? {A} {Review} of the {Research}.
\newblock {\em Journal of Engineering Education}, 93(3):223--231.
\newblock \_eprint:
  https://onlinelibrary.wiley.com/doi/pdf/10.1002/j.2168-9830.2004.tb00809.x.

\bibitem[Rivers and Koedinger, 2017]{rivers_data-driven_2017}
Rivers, K. and Koedinger, K.~R. (2017).
\newblock Data-{Driven} {Hint} {Generation} in {Vast} {Solution} {Spaces}: a
  {Self}-{Improving} {Python} {Programming} {Tutor}.
\newblock {\em International Journal of Artificial Intelligence in Education},
  27(1):37--64.

\bibitem[Rogers et~al., 2014]{rogers_acce_2014}
Rogers, S., Tang, S., and Canny, J. (2014).
\newblock {ACCE}: automatic coding composition evaluator.
\newblock In {\em Proceedings of the first {ACM} conference on {Learning} @
  scale conference}, L@{S} '14, pages 191--192, New York, NY, USA. Association
  for Computing Machinery.

\bibitem[Romero and Ventura, 2010]{romero_educational_2010}
Romero, C. and Ventura, S. (2010).
\newblock Educational {Data} {Mining}: {A} {Review} of the {State} of the
  {Art}.
\newblock {\em IEEE Transactions on Systems, Man, and Cybernetics, Part C
  (Applications and Reviews)}, 40(6):601--618.
\newblock Conference Name: IEEE Transactions on Systems, Man, and Cybernetics,
  Part C (Applications and Reviews).

\bibitem[R{\"o}{\ss}ling et~al., 2008]{rosling_enhancing_2008}
R{\"o}{\ss}ling, G., Joy, M., Moreno, A., Radenski, A., Malmi, L., Kerren, A.,
  Naps, T., Ross, R.~J., Clancy, M., Korhonen, A., Oechsle, R., and Iturbide,
  J. {\'A}.~V. (2008).
\newblock Enhancing learning management systems to better support computer
  science education.
\newblock {\em ACM SIGCSE Bulletin}, 40(4):142--166.

\bibitem[Sakimura et~al., 2014]{sakimura_openid_2014}
Sakimura, N., Bradley, J., Jones, M., De~Medeiros, B., and Mortimore, C.
  (2014).
\newblock Openid connect core 1.0.
\newblock {\em The OpenID Foundation}, page~S3.

\bibitem[Schleimer et~al., 2003]{schleimer_winnowing_2003}
Schleimer, S., Wilkerson, D.~S., and Aiken, A. (2003).
\newblock Winnowing: local algorithms for document fingerprinting.
\newblock In {\em Proceedings of the 2003 {ACM} {SIGMOD} international
  conference on {Management} of data}, {SIGMOD} '03, pages 76--85, New York,
  NY, USA. Association for Computing Machinery.

\bibitem[Schunk and Zimmerman, 1994]{schunk_self-regulation_1994}
Schunk, D.~H. and Zimmerman, B.~J. (1994).
\newblock {\em Self-regulation of learning and performance: {Issues} and
  educational applications.}
\newblock Lawrence Erlbaum Associates, Inc.

\bibitem[Staubitz et~al., 2015]{staubitz_towards_2015}
Staubitz, T., Klement, H., Renz, J., Teusner, R., and Meinel, C. (2015).
\newblock Towards practical programming exercises and automated assessment in
  {Massive} {Open} {Online} {Courses}.
\newblock In {\em 2015 {IEEE} {International} {Conference} on {Teaching},
  {Assessment}, and {Learning} for {Engineering} ({TALE})}, pages 23--30.

\bibitem[Stenerson and Dawson, 1998]{stenerson_internet_1998}
Stenerson, D. and Dawson, F. (1998).
\newblock Internet {Calendaring} and {Scheduling} {Core} {Object}
  {Specification} ({iCalendar}).
\newblock Request for {Comments} RFC 2445, Internet Engineering Task Force.
\newblock Num Pages: 148.

\bibitem[Stevens et~al., 1999]{stevens_structured_1999}
Stevens, W.~P., Myers, G.~J., and Constantive, L.~L. (1999).
\newblock Structured design.
\newblock {\em IBM Systems Journal}, 38(2.3):231--256.
\newblock Conference Name: IBM Systems Journal.

\bibitem[Strijbol et~al., 2022]{strijbol_tested_nodate}
Strijbol, N., Van~Petegem, C., Maertens, R., Sels, B., Scholliers, C., Dawyndt,
  P., and Mesuere, B. (2022).
\newblock {TESTed} {\textemdash} an educational test framework for
  programming-language-agnostic exercises.
\newblock {\em IEEE Transactions on Learning Technologies}.

\bibitem[Tucker, 2012]{tucker_flipped_2012}
Tucker, B. (2012).
\newblock The flipped classroom.
\newblock {\em Education next}, 12(1):82--83.

\bibitem[Tuomi, 2013]{tuomi_open_2013}
Tuomi, I. (2013).
\newblock Open {Educational} {Resources} and the {Transformation} of
  {Education}.
\newblock {\em European Journal of Education}, 48(1):58--78.
\newblock \_eprint: https://onlinelibrary.wiley.com/doi/pdf/10.1111/ejed.12019.

\bibitem[Van~Petegem et~al., 2022]{van_petegem_passfail_2022}
Van~Petegem, C., Deconinck, L., Mourisse, D., Maertens, R., Strijbol, N.,
  Dhoedt, B., De~Wever, B., Dawyndt, P., and Mesuere, B. (2022).
\newblock Pass/{Fail} {Prediction} in {Programming} {Courses}.
\newblock {\em Journal of Educational Computing Research}, page
  07356331221085595.
\newblock Publisher: SAGE Publications Inc.

\bibitem[Verhoeff, 2008]{verhoeff_programming_2008}
Verhoeff, T. (2008).
\newblock Programming {Task} {Packages}: {Peach} {Exchange}.
\newblock {\em Olympiads in Informatics}, page 192.
\newblock Publisher: Citeseer.

\bibitem[Wasik et~al., 2018]{wasik_survey_2018}
Wasik, S., Antczak, M., Badura, J., Laskowski, A., and Sternal, T. (2018).
\newblock A {Survey} on {Online} {Judge} {Systems} and {Their} {Applications}.
\newblock {\em ACM Computing Surveys}, 51(1):3:1--3:34.

\bibitem[Wiegers, 1996]{wiegers_creating_1996}
Wiegers, K.~E. (1996).
\newblock {\em Creating a software engineering culture}.
\newblock Pearson Education.

\bibitem[Wilcox, 2016]{wilcox_testing_2016}
Wilcox, C. (2016).
\newblock Testing {Strategies} for the {Automated} {Grading} of {Student}
  {Programs}.
\newblock In {\em Proceedings of the 47th {ACM} {Technical} {Symposium} on
  {Computing} {Science} {Education}}, {SIGCSE} '16, pages 437--442, New York,
  NY, USA. Association for Computing Machinery.

\bibitem[Wiley et~al., 2014]{wiley_open_2014}
Wiley, D., Bliss, T.~J., and McEwen, M. (2014).
\newblock Open {Educational} {Resources}: {A} {Review} of the {Literature}.
\newblock In Spector, J.~M., Merrill, M.~D., Elen, J., and Bishop, M.~J.,
  editors, {\em Handbook of {Research} on {Educational} {Communications} and
  {Technology}}, pages 781--789. Springer, New York, NY.

\bibitem[Wilkinson et~al., 2016]{wilkinson_fair_2016}
Wilkinson, M.~D., Dumontier, M., Aalbersberg, I.~J., Appleton, G., Axton, M.,
  Baak, A., Blomberg, N., Boiten, J.-W., da~Silva~Santos, L.~B., Bourne, P.~E.,
  Bouwman, J., Brookes, A.~J., Clark, T., Crosas, M., Dillo, I., Dumon, O.,
  Edmunds, S., Evelo, C.~T., Finkers, R., Gonzalez-Beltran, A., Gray, A. J.~G.,
  Groth, P., Goble, C., Grethe, J.~S., Heringa, J., {\textquoteright}t~Hoen, P.
  A.~C., Hooft, R., Kuhn, T., Kok, R., Kok, J., Lusher, S.~J., Martone, M.~E.,
  Mons, A., Packer, A.~L., Persson, B., Rocca-Serra, P., Roos, M., van Schaik,
  R., Sansone, S.-A., Schultes, E., Sengstag, T., Slater, T., Strawn, G.,
  Swertz, M.~A., Thompson, M., van~der Lei, J., van Mulligen, E., Velterop, J.,
  Waagmeester, A., Wittenburg, P., Wolstencroft, K., Zhao, J., and Mons, B.
  (2016).
\newblock The {FAIR} {Guiding} {Principles} for scientific data management and
  stewardship.
\newblock {\em Scientific Data}, 3(1):160018.
\newblock Bandiera\_abtest: a Cg\_type: Nature Research Journals Number: 1
  Primary\_atype: Comments \& Opinion Publisher: Nature Publishing Group
  Subject\_term: Publication characteristics;Research data Subject\_term\_id:
  publication-characteristics;research-data.

\bibitem[Williams et~al., 2002]{williams_support_2002}
Williams, L., Wiebe, E., Yang, K., Ferzli, M., and Miller, C. (2002).
\newblock In {Support} of {Pair} {Programming} in the {Introductory} {Computer}
  {Science} {Course}.
\newblock {\em Computer Science Education}, 12(3):197--212.
\newblock Publisher: Routledge \_eprint:
  https://doi.org/10.1076/csed.12.3.197.8618.

\bibitem[Wing, 2006]{wing_computational_2006}
Wing, J.~M. (2006).
\newblock Computational thinking.
\newblock {\em Communications of the ACM}, 49(3):33--35.
\newblock Publisher: ACM New York, NY, USA.

\bibitem[Woit and Mason, 2003]{woit_effectiveness_2003}
Woit, D. and Mason, D. (2003).
\newblock Effectiveness of online assessment.
\newblock In {\em Proceedings of the 34th {SIGCSE} technical symposium on
  {Computer} science education}, {SIGCSE} '03, pages 137--141, New York, NY,
  USA. Association for Computing Machinery.

\bibitem[Wootton, 2002]{wootton_encouraging_2002}
Wootton, S. (2002).
\newblock Encouraging {Learning} or {Measuring} {Failure}?
\newblock {\em Teaching in Higher Education}, 7(3):353--357.
\newblock Publisher: Routledge \_eprint:
  https://doi.org/10.1080/13562510220144833.

\bibitem[Yourdon and Constantine, 1979]{yourdon_structured_1979}
Yourdon, E. and Constantine, L.~L. (1979).
\newblock Structured design. {Fundamentals} of a discipline of computer program
  and systems design.
\newblock {\em Englewood Cliffs: Yourdon Press}.

\end{thebibliography}

\end{document}